\documentclass[secnumarabic,amssymb,nobibnotes,aps,prb,preprint,letterpaper,preprintnumbers,amsmath,superscriptaddress,endfloats*]{revtex4-2}
\usepackage{graphicx}
\usepackage{dcolumn}
\usepackage{bm}
\usepackage{amsmath,amssymb,bm}
\usepackage{graphicx}
\usepackage{xspace}
\usepackage{latexsym}
\usepackage{color}
\usepackage{xr}
\usepackage{siunitx}

\usepackage{lineno}

\usepackage[colorlinks = true,linkcolor = blue,urlcolor  = red,citecolor = red,anchorcolor = blue]{hyperref}
\usepackage{placeins}
\usepackage{orcidlink}
\usepackage{glossaries}
\usepackage{cleveref}
\usepackage{gensymb}
\usepackage{zref-user,zref-xr}

\newacronym{INS}{INS}{inelastic neutron scattering}
\newacronym{ARPES}{ARPES}{angular resolved photo emission spectroscopy}
\newacronym{RIXS}{RIXS}{resonant inelastic X-ray scattering}
\newacronym{OPRIXS}{op-RIXS}{operando resonant inelastic X-ray scattering}
\newacronym{QSL}{QSL}{quantum spin liquid}
\newacronym{DSF}{DSF}{dynamical structure factor}
\newacronym{IMT}{IMT}{insulator-to-metal transition}
\newacronym{XAS}{XAS}{X-ray absorption spectroscopy}
\newacronym{SEM}{SEM}{scanning electron microscopy}
\newacronym{IV}{$I-V$}{current-voltage}
\newacronym{80meVMode}{spin-orbital}{spin-orbital}
\newacronym{NDR}{negative differential resistance}{negative differential resistance}
\newacronym{NESS}{non-equilibrium steady state}{non-equilibrium steady state}
\newacronym{QCP}{QCP}{quantum critical point}
\newacronym{R}{$R$}{resistance}
\newacronym{TMDC}{2D-TMDC}{two-dimensional transition-metal dichalcogenide} 

\begin{document}



\title{\textbf{Evidence of electronic states driving current-induced insulator-to-metal transition }}
\author{V. K. Bhartiya}
\email{vbhartiya1@bnl.gov}
\affiliation{National Synchrotron Light Source II, Brookhaven National Laboratory, Upton, New York 11973, USA}
\author{R. Hartmann}
\affiliation{Department of Physics, University of Konstanz, 78457 Konstanz, Germany}
\author{F. Forte}
\email{filomena.forte@spin.cnr.it}
\affiliation{CNR-SPIN, c/o Universit\`a di Salerno, I-84084 Fisciano, Salerno, Italy}
\author{F. Gabriele}
\affiliation{CNR-SPIN, c/o Universit\`a di Salerno, I-84084 Fisciano, Salerno, Italy}
\author{T. Kim}
\affiliation{National Synchrotron Light Source II, Brookhaven National Laboratory, Upton, New York 11973, USA}
\author{G. Cuono}
\affiliation{CNR-SPIN, c/o Universit\`a “G. d\textquotesingle Annunzio”, I-66100 Chieti, Italy }
\author{C. Autieri}
\affiliation{International Research Centre Magtop, Institute of Physics, Polish Academy of Sciences, Aleja Lotnik\'ow 32/46, 02668 Warsaw, Poland}
\author{S. Fan}
\affiliation{National Synchrotron Light Source II, Brookhaven National Laboratory, Upton, New York 11973, USA}

\author{K. Kisslinger}
\affiliation{Center for Functional Nanomaterials, Brookhaven National Laboratory, Upton, New York 11973, USA }

\author{F. Camino}
\affiliation{Center for Functional Nanomaterials, Brookhaven National Laboratory, Upton, New York 11973, USA }

\author{M. Lettieri}
\affiliation{CNR-SPIN, c/o Universit\`a di Salerno, I-84084 Fisciano, Salerno, Italy}

\author{R. Fittipaldi}
\affiliation{CNR-SPIN, c/o Universit\`a di Salerno, I-84084 Fisciano, Salerno, Italy}

\author{C. Mazzoli}
\affiliation{National Synchrotron Light Source II, Brookhaven National Laboratory, Upton, New York 11973, USA}

\author{D. N. Basov}
\affiliation{Department of Physics, Columbia University, New York, New York 10027, USA.}

\author{J. Pelliciari}
\affiliation{National Synchrotron Light Source II, Brookhaven National Laboratory, Upton, New York 11973, USA}

\author{A. Di Bernardo}
\affiliation{Department of Physics, University of Konstanz, 78457 Konstanz, Germany}
\affiliation{Universit\`a di Salerno, Department of Physics ``E. R. Caianiello", 84084 Fisciano, Salerno, Italy}

\author{A. Vecchione}
\affiliation{CNR-SPIN, c/o Universit\`a di Salerno, I-84084 Fisciano, Salerno, Italy}
\author{M. Cuoco}
\affiliation{CNR-SPIN, c/o Universit\`a di Salerno, I-84084 Fisciano, Salerno, Italy}
\author{V. Bisogni}
\email{bisogni@bnl.gov}
\affiliation{National Synchrotron Light Source II, Brookhaven National Laboratory, Upton, New York 11973, USA}

\date{\today}
\maketitle


\textbf{On demand current-driven  \gls*{IMT} is pivotal for the next generation of energy-efficient and scalable microelectronics. 
IMT is a key phenomenon observed in various quantum materials, and it is enabled by the complex interplay of spin, lattice, charge, and orbital degrees of freedom (DOF). Despite significant prior work, the underlying mechanism of the current-driven \gls*{IMT} remains elusive, primarily due to the difficulty in simultaneously 
obtaining bulk fingerprints of all the electronic DOF. 
Here, we employ in-operando resonant inelastic x-ray scattering (RIXS) on Ca$_2$RuO$_4$, a prototypical strongly correlated material, to track the evolution of the electronic DOF encoded in the RIXS spectra during the current-driven \gls*{IMT}. Upon entering the conductive state, we observe an energy-selective suppression of the RIXS intensity, proportional to the current. 
Using complementary RIXS cross-section calculations, we demonstrate that the non-equilibrium conductive state 
emerges from the formation of correlated electronic states with a persistent Mott gap.} 

\begin{figure*}
    \centering
    \includegraphics[width=0.95\linewidth]{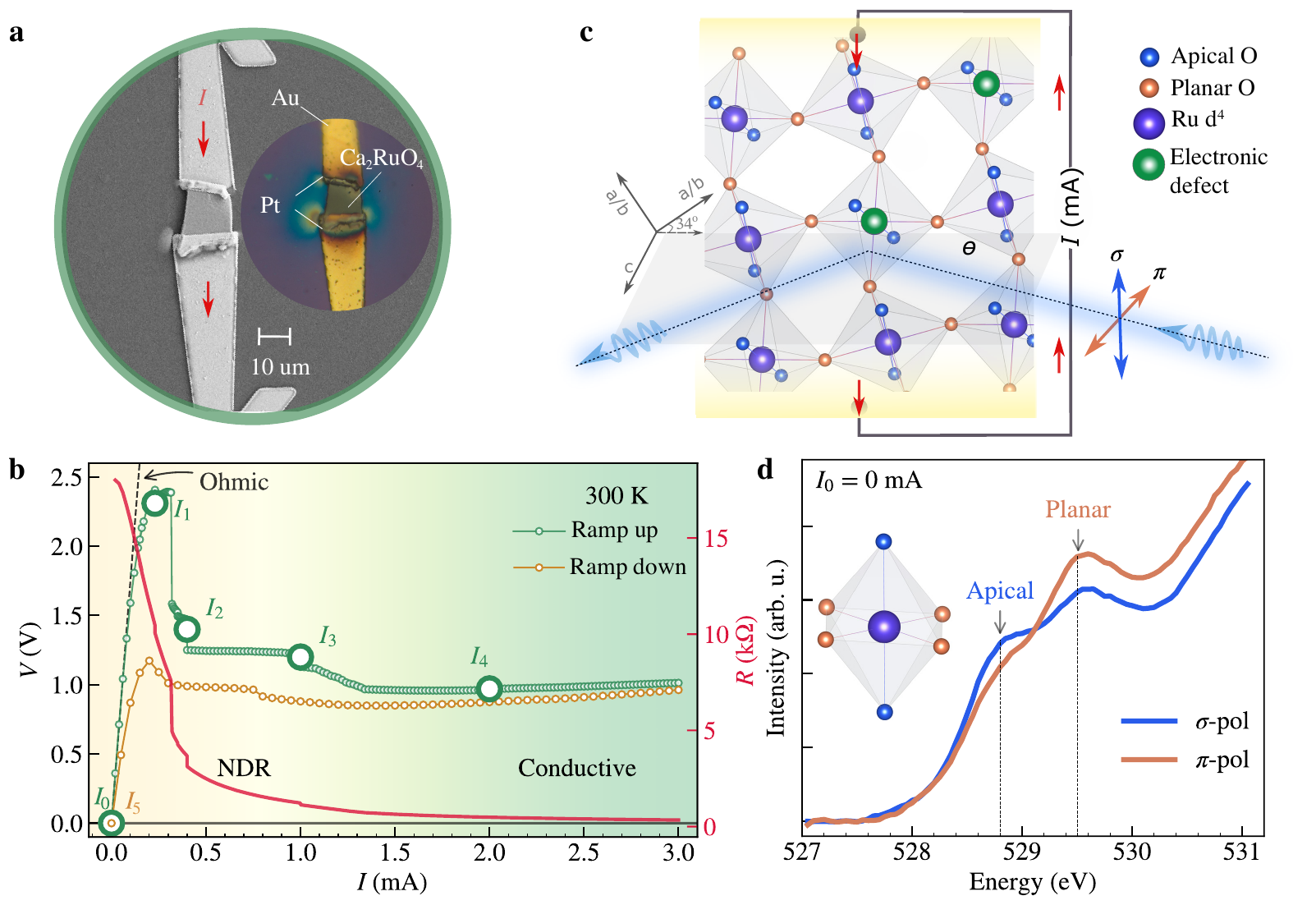}
    \caption{\textbf{Ca$_2$RuO$_4$ device, experimental geometry, \gls*{IV} characteristics, and XAS}. 
     \textbf{a}, \gls*{SEM} (gray-scale) and optical (color-scale) images of the Ca$_{2}$RuO$_4$ device used for in-operando \gls*{RIXS}. 
    \textbf{b},  
    \gls*{IV} curve (dotted green/brown for 
    d.c. current ramping up/down) measured during in-operando \gls*{RIXS}, at 300 K. The circles $I_n$ represent the in-operando \gls*{RIXS} measurement points, i.e. $I_0$ to $I_4$ are measured for increasing current, while $I_5$=0 mA is measured right after ramping down the current. The red curve is the electrical resistance \gls*{R}. The dashed line highlights the linear Ohmic trend expected at low current. Above 0.2$-$0.3 mA, a \gls*{NDR} (NDR) state appears and eventually at higher current a conductive \gls*{NESS} is stabilized.
    \textbf{c}, XAS and in-operando \gls*{RIXS} experimental  geometry. The RIXS scattering plane includes the sample \textit{c}-axis, while the in-plane \textit{a/b}-axis is off by $\sim$ 34$^{\circ}$.  Axes \textit{a} and \textit{b} could not be resolved, thus represented as \textit{a/b}-axis. The x-ray beam impinges on the sample at an angle $\theta$ = 45$^{\circ}$, and it is linearly polarized ($\sigma$ or $\pi$). The Ca$_{2}$RuO$_4$ crystal structure is represented in the background, with the Ru atoms ($d^4$, blue) contained in octahedral oxygen cages. The current injected by the electrodes creates correlated electronic defects on the Ru sites (green spheres). 
    \textbf{d}, O $K-$edge \gls*{XAS} of the Ca$_{2}$RuO$_4$ device at $I_0$ = 0 mA, 300 K. 
    The $\sigma$ ($\pi$) spectrum resonates mostly with the apical (planar) oxygens.}
    \label{fig:Fig1}
\end{figure*}

\section{Main}

The insulator-to-metal transition (IMT) is a fundamental physical phenomenon that refers to the change from a non-conductive to a conductive state \cite{Mott1968RMP, Imada1998RMP}, and it is observed in a variety of quantum materials, such as strongly correlated oxides and transition-metal dichalcogenides ~\cite{Alexander1999PRB,Kalcheim2020NatComm,Radisavljevic2013NatMat,YasuiPRB2024,Shao2018NAM,SarmaPRB2003}. The \gls*{IMT} is typically explained by the closure of a gap in the energy spectrum \cite{Sutter2017HallmarksCa2RuO4,YasuiPRB2024,Shao2018NAM}. This mechanism underlies \gls*{IMT}s induced by temperature, pressure, strain, or doping, all of which lead to the formation of an equilibrium metallic phase ~\cite{Zylbersztejn1975PRB,Chen2017NanoLett,PARIJA20201166, Lee2018Science,Alexander1999PRB,Nakamura2013Electricfield, Sindhu2023NatComm, Nakatsuji2000doping,Ricco2018strain}. 
Additionally, a 
conductive \gls*{NESS} can be triggered by external stimuli, like ultra-fast laser pumps, electric fields or direct (d.c.) currents \cite{Nakamura2013Electricfield, Bertinshaw2020UniqueState, Crunteanu2010,Zhang2016ACSNano,Cirillo2019EmergenceCa2RuO4,Sindhu2023NatComm,Verma2024NatPhy,Curcio2023Current-drivenCa2RuO4,Suen2023NatureTransport-ARPES,Torre2021_RevModPhys,Javier2021Science,Xinwei2025Nature}. 
The current-induced \gls*{IMT}s are of particular interest due to their potential for on-demand conductivity, rapid switching, and easily scalable setups, making them ideal for low-power electronics and neuromorphic computing ~\cite{Zhou2015IEEE,Park2023AdM}. Consequently, a current-induced conductive \gls*{NESS} represents a compelling opportunity from both a fundamental and a technological standpoint. However, to fully leverage this phenomenon, it is crucial to elucidate the mechanisms governing its formation and its distinctions from, or similarities to, the equilibrium metallic state.

The strongly correlated Mott-insulator Ca$_{2}$RuO$_4$ is a prototypical compound for \gls*{IMT} ($T_{\rm{IMT}} \approx$ 357 K \cite{Zylbersztejn1975PRB}), exhibiting a complex ground state with intertwined electronic DOF (e.g. spin, lattice, orbital and charge) \cite{Sutter2017HallmarksCa2RuO4, Das2018Spin-OrbitalScattering, Nakamura2023UniqueItinerant, Jain2017HiggsAntiferromagnet} and a rich phase diagram \cite{Alexander1999PRB,  Nakatsuji2000doping, Ricco2018strain, Friedt01}. Notably, Ca$_{2}$RuO$_4$ demonstrate a conductive \gls*{NESS} driven by a small current, even at room temperature \cite{Nakamura2013Electricfield, IchiroJSPS2020, Bertinshaw2020UniqueState, Cirillo2019EmergenceCa2RuO4,Curcio2023Current-drivenCa2RuO4,Suen2023NatureTransport-ARPES, Zhang_PRX_2019}. Various experimental techniques 
have been used so far to characterize the current-driven \gls*{IMT} in Ca$_{2}$RuO$_4$~\cite{Nakamura2013Electricfield, Bertinshaw2020UniqueState, Crunteanu2010,Curcio2023Current-drivenCa2RuO4,Fursich2019RamanCa2RuO4}, however, a comprehensive understanding of the conductive \gls*{NESS} remains elusive. 
Specifically, while the equilibrium metallic state and the accompanying melting of the Mott gap are triggered by the underlying structural transition \cite{Alexander1999PRB,Bertinshaw2020UniqueState}, it has been observed that the current-driven \gls*{NESS} is associated with a different structural rearrangement
~\cite{Bertinshaw2020UniqueState, Nakamura2013Electricfield}. 
Furthermore, recent electronic structural investigations by angular-resolved photoemission spectroscopy (ARPES) reveal the persistence of the Mott gap in the current-driven \gls*{NESS}, suggesting an electronic origin distinct from the equilibrium metallic state \cite{Curcio2023Current-drivenCa2RuO4, Suen2023NatureTransport-ARPES}. 
Therefore, to fully uncover the underlying mechanism of the conductive \gls*{NESS}, it is essential to simultaneously probe all the electronic DOF with bulk sensitivity, capturing their 
interactions and their role during the current-driven \gls*{IMT}. In-operando \gls*{RIXS} at the O $K$-edge, with its inherent bulk sensitivity and ability to directly probe  excitations associated with all the electronic DOF,  offers a unique opportunity to address this challenge. 

In this work, we employed in-operando \gls*{RIXS} to investigate the electronic DOF in Ca$_{2}$RuO$_4$ as a function of d.c. current, while stabilizing the conductive \gls*{NESS}, providing a comprehensive understanding of its electronic nature.
The O $K$-edge RIXS was previously used to identify 
the Ru$^{4+}$(4$d^4$) spin-orbital entangled insulating ground state of Ca$_{2}$RuO$_4$, and the resulting excitations are well understood in terms of, e.g., spin-orbital excitations, spin-orbit exciton, Hund's modes, and $dd$-excitations ~\cite{Fatuzzo2015Spin-orbit-inducedStudy, Das2018Spin-OrbitalScattering,Arx2023}.
Here, we identify an energy selective suppression of the RIXS intensity as we supply d.c. current, and attribute this to the changeover from an insulating state to conductive \gls*{NESS}. 
Complementary investigations of the thermally-induced 
metallic phase by RIXS demonstrate that the current-driven conductive \gls*{NESS} is electronically distinct.
RIXS cross-section calculations reproduce the observed changes as a function of d.c. current bias by introducing correlated electronic states (e.g. $d^3$) to the $d^4$ insulating ground state configuration. Such extra states can drive the current-induced \gls*{IMT} with persistent Mott gap.
\begin{figure*}
    \centering
    \includegraphics[width=1\linewidth]{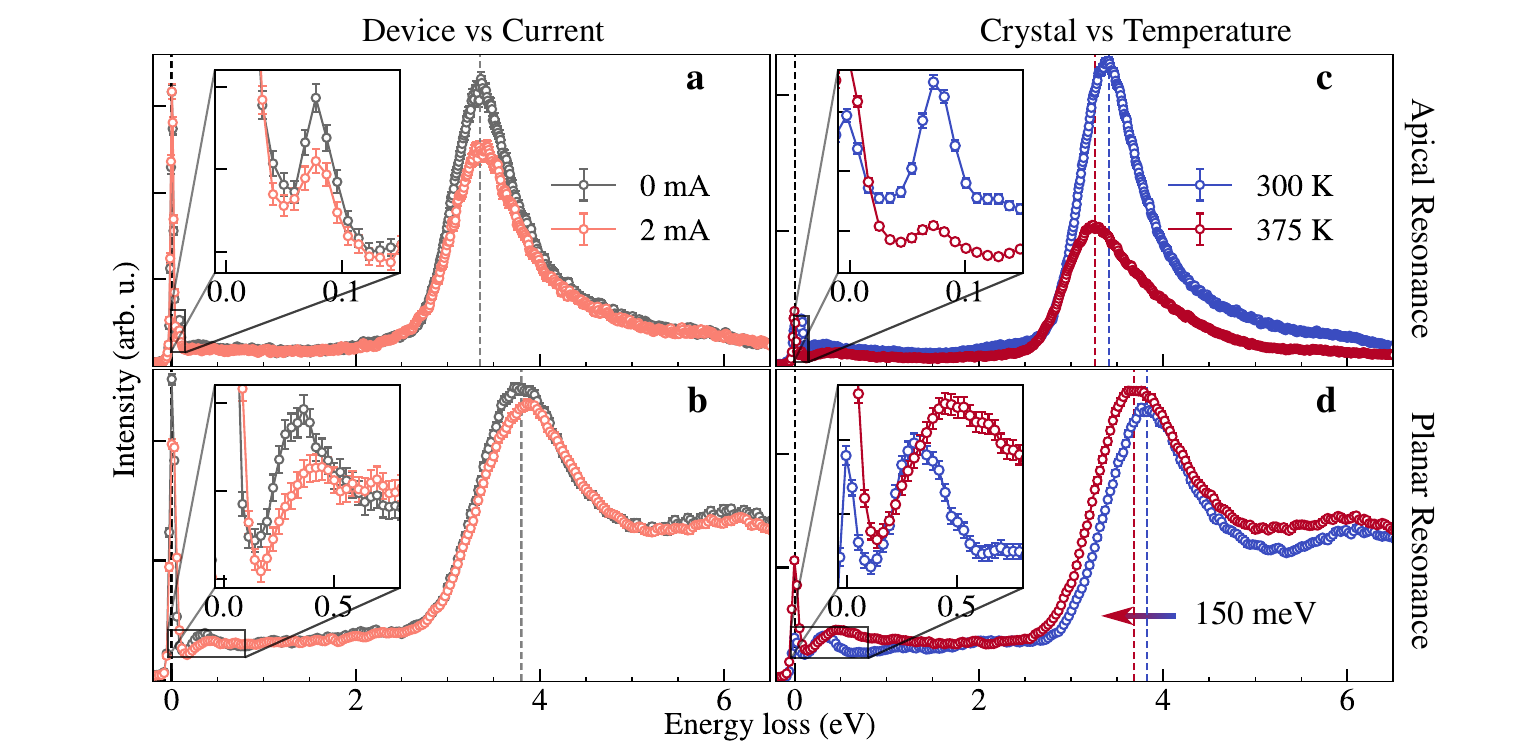}
    \caption{\textbf{Effect of current-driven and temperature-driven IMT in RIXS spectra}. {\bf a-b}, O $K-$edge RIXS spectrum in the insulating ground state (0 mA, gray curve) and in the current-induced conductive state (2 mA, orange curve), measured at the apical resonance ({\bf a}), and at the planar resonance ({\bf b}), at 300 K.  {\bf c-d}, Temperature dependent O $K-$edge RIXS spectrum  below (300 K, blue curve) and above (375 K, red curve) thermally-induced \gls*{IMT} ($T_\text{IMT}$ = 357 K) measured at the apical resonance ({\bf c}), and at the planar resonance ({\bf d}). Figures inset highlight evolution of the low energy  $\approx$ 80 meV  and $\approx$ 350 meV spectral weight. Dotted vertical lines help tracking the softening ($\approx$ 150 meV, blue to red arrow) observed in the $dd$-peak for the temperature-driven \gls*{IMT}.}
    \label{fig:Fig2}
\end{figure*}
\section{Spectroscopic evidence of current-induced conductive state}

To perform in-operando \gls*{RIXS} experiment, we fabricated a device with a micrometer-sized flake (10~\SI{}{\micro\metre} $\times$ 12~\SI{}{\micro\metre} laterally, 1~\SI{}{\micro\metre} thick) exfoliated from a Ca$_{2}$RuO$_4$ single crystal. The flake was selected on a SiO$_2$/Si substrate, and Au/Ti contacts were deposited on its surface to realize a two-point configuration, see Fig. \hyperref[fig:Fig1]{\ref*{fig:Fig1}a}. 
The micrometer-size of the Ca$_{2}$RuO$_4$ device  
was preferred for the following reasons. First, it ensures that the majority of the sample is probed by in-operando \gls*{RIXS} (beamspot  10  $\times$ 3 \SI{}{\micro\metre}$^2$) while undergoing the \gls*{IMT}, as the transition to the conductive state is known to be spatially inhomogeneous ~\cite{Zhang_PRX_2019,Gauquelin2023NanoLett}. Second, small samples are more likely to be free from structural defects and 
vacancies that are detrimental to both transport and mechanical stability, while big crystals are prone to shattering under long or multiple operations. 
Figure \hyperref[fig:Fig1]{\ref*{fig:Fig1}b} presents the \gls*{IV} curve of the device collected during the in-operando \gls*{RIXS} experiment, with the green (grey) line representing the \gls*{IV} profile measured for increasing (decreasing) d.c. current bias. For current values $\gtrsim 0.2$ mA, a distinctive \gls*{NDR}~\cite{Bradicich2023AEM} is observed, providing a direct signature of the current-driven \gls*{IMT} \cite{Nakamura2013Electricfield,Cirillo2019EmergenceCa2RuO4}. The electrical resistance concomitantly drops by three orders of magnitude as the current increases. More details about sample, device fabrication and characterization are provided in the Methods and Supplementary Section  S1. 

To investigate the ground state evolution of Ca$_{2}$RuO$_4$ across its current-driven \gls*{IMT}, we performed in-operando \gls*{RIXS} at selected d.c. current values ($I_0$, $I_1$, $I_2$, $I_3$, and $I_4$), highlighted with large open green circles in Fig. \hyperref[fig:Fig1]{\ref*{fig:Fig1}b}. To stabilize the \gls*{NESS}, the device was kept in constant d.c. current flow during the in-operando \gls*{RIXS} measurements~\cite{Nakamura2013Electricfield,IchiroJSPS2020,Nakamura2023UniqueItinerant,Bertinshaw2020UniqueState} (see Supplementary Fig. S1 and Section S2.1). The scattering geometry is represented in Fig. \hyperref[fig:Fig1]{\ref*{fig:Fig1}c}. The Ca$_{2}$RuO$_4$ $c$-axis lied in the scattering plane, while the in-plane axes were determined to be 34$^{\circ}$ off azimuthally by scanning electron microscopy (SEM), see Supplementary S1.2. The electrodes were 
oriented orthogonally to the scattering plane. All measurements were performed at 300 K and at fixed geometry, see Methods. \gls*{XAS} was used to identify the resonant energies for the in-operando \gls*{RIXS} measurements. Figure \hyperref[fig:Fig1]{\ref*{fig:Fig1}d} presents the O $K$-edge \gls*{XAS} measured in the insulating ground state Ru$^{4+}$(4$d^4$), without current bias. The two peaks observed at 528.8 eV and 529.5 eV correspond to the transition from 1$s$ to the 2$p$ orbitals hybridized with the Ru $t_\text{2g}$ apical and planar orbitals, respectively ~\cite{Fatuzzo2015Spin-orbit-inducedStudy,Das2018Spin-OrbitalScattering}. The apical (planar) resonance is enhanced by $\sigma$ ($\pi$) incident polarization. Since these two resonances provide access to different low-energy excitations in the RIXS spectra~\cite{Das2018Spin-OrbitalScattering}, we measured O $K$-edge RIXS spectra for both apical ($\sigma$ polarization) and planar ($\pi$ polarization) resonances.  


Figure \hyperref[fig:Fig2]{\ref*{fig:Fig2}} presents an overview of the RIXS spectral changes observed across the current-driven \gls*{IMT} with respect to a thermal-driven \gls*{IMT}, to assess the distinctiveness or similarities of these two stimuli and the possibility of Joule heating under d.c. current bias.  Figure \hyperref[fig:Fig2]{\ref*{fig:Fig2}a} (\hyperref[fig:Fig2]{\ref*{fig:Fig2}b}) presents the \gls*{RIXS} spectra at apical (planar) resonance of the CRO device in the insulating ground state ($I_0$ = 0 mA) and in the current-driven conductive state ($I_4$ = 2 mA). With similar layout, Figs. \hyperref[fig:Fig2]{\ref*{fig:Fig2}c-d} present the RIXS spectra of a reference CRO single crystal in the insulating ground state ($T$ = 300 K) and in the metallic state ($T$ = 375 K).
In both the current- and temperature- driven cases, the RIXS spectra display changes across the \gls*{IMT}, suggesting a reorganization of the electronic DOF, e.g. spin, orbital, charge and lattice, that are involved in the probed excitations as discussed in the following. Furthermore, the observed spectral changes are distinct for the two stimuli.
The evolution of the RIXS spectral weight across the current-driven IMT mostly displays an intensity suppression at both apical and planar resonances. However, such a suppression is not a rescaling of the overall spectrum, but appears as ``energy-selective'', i.e. distinct suppression rate for each excitation. No significant peak shift or line shape change is identified. Contrary, the evolution of the RIXS spectral weight across the thermal-driven IMT displays both intensity suppression and enhancement, depending on the excitation of interest. Relevant is the increase of spectral weight and line shape evolution of the $\approx$ 350 meV excitation (see Fig. \hyperref[fig:Fig2]{\ref*{fig:Fig2}d}). Furthermore, a $\approx$ 150 meV softening of the high-energy ($>$ 3 eV) excitations is observed in the metallic phase. 

Overall, the spectral differences observed under current and temperature suggest that these two stimuli induce electronically 
distinct conductive states, and consequently, that the current-driven \gls*{IMT} is not caused by Joule heating effects
~\cite{Alexander1999PRB}. This is further confirmed by Supplementary Figure S3, where the \gls*{RIXS} spectrum measured right after switching the d.c. current bias off ($I_5$ = 0 mA) displays a full recovery of the spectral weight acquired before activating the device ($I_0$ = 0 mA).
While the character of the  excitations and the spectral weight evolution observed across the current-driven \gls*{IMT} 
will be discussed later, here we note that the low-energy spectral weight (< 1 eV) measured in the thermal-induced metallic phase (see inset of Fig. \hyperref[fig:Fig2]{\ref*{fig:Fig2}d}) is consistent with the particle-hole continuum observed by RIXS in other metallic ruthenates ~\cite{Fatuzzo2015Spin-orbit-inducedStudy,KimPRL2012}, thus suggestive of the closure of the Mott insulating band-gap ($\approx -$ 0.4 eV for Ca$_{2}$RuO$_4$~\cite{Sutter2017HallmarksCa2RuO4}). Such a low-energy spectral weight redistribution is absent in the spectrum under d.c. current bias (see inset of Fig. \hyperref[fig:Fig2]{\ref*{fig:Fig2}b}), supporting the persistence of the Mott insulating band-gap in the conductive \gls*{NESS} \cite{Curcio2023Current-drivenCa2RuO4,Suen2023NatureTransport-ARPES}. 

\begin{figure*}
    \centering
    \includegraphics[width=1\linewidth]{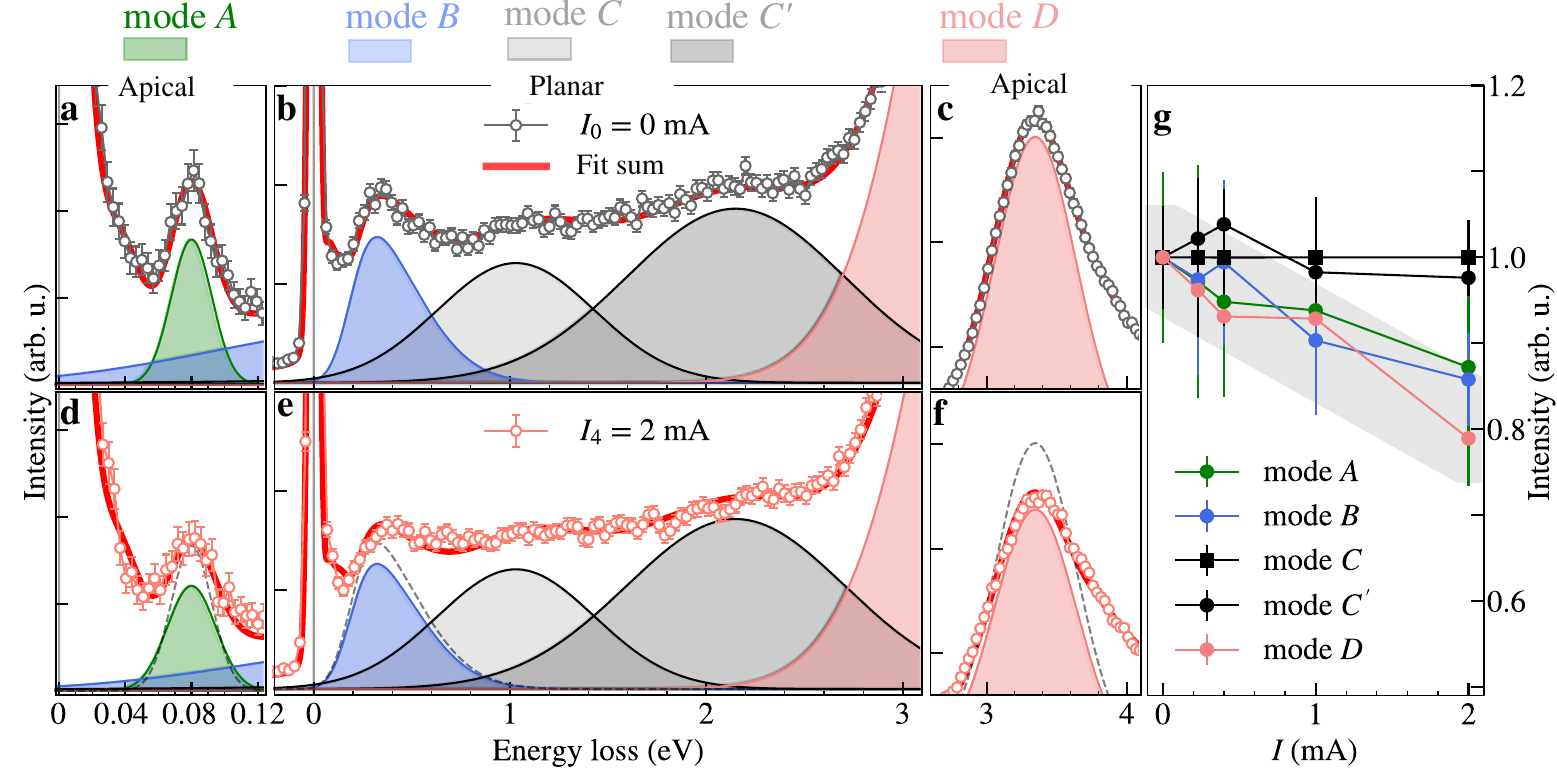}
    \caption{\textbf{RIXS spectra versus current}. {\bf a}--{\bf c}, Fitted RIXS spectrum for the insulating ground state (0 mA) and, {\bf d}--{\bf f},  for the current-driven  conductive state (2 mA), at 300 K. Shaded regions and bold lines indicate fitted modes. Dotted lines in ({\bf d}–{\bf f}) show 0 mA fitted modes for comparison.  The shaded regions (green$-$ mode \textit{A}, blue$-$ mode \textit{B}, gray$-$ mode \textit{C} $\&$ \textit{C}$^{\prime}$, and red$-$ mode \textit{D}) are various characteristic excitations of Ca$_{2}$RuO$_4$ and highlight the energy-selective changes in the RIXS spectrum. Modes \textit{A} and  \textit{D} are measured  at the apical resonant and modes \textit{B}, \textit{C}, and \textit{C}$^{\prime}$ at the planar resonance. The spectral weight associated with mode \textit{C} $\&$ \textit{C}$^{\prime}$ centered at $\approx$ 1 eV and 2 eV do not evolve with the current. {\bf g},  Evolution of the fitted RIXS modes {\it vs} current. The data points are the integrated spectral weight of the respective fitted modes shown in panels (\textbf{a}-\textbf{f}), normalized to their weight at $I_0$ = 0 mA. The gray colored background highlights a linear suppression of RIXS spectral weight with current. The error bars are one standard deviation. 
    }
    \label{fig:Fig3}
\end{figure*}

Figure \hyperref[fig:Fig3]{\ref*{fig:Fig3}} displays the evolution of the \gls*{RIXS} spectral weight as a function of d.c. current bias, from 0~mA to 2~mA. We identify several excitations involving all the aforementioned electronic DOF, labeled as \textit{A}, \textit{B}, \textit{C}, \textit{C}$^\prime$, and \textit{D}, and interpret them on the basis of the multiplet structure of the Ru$^{4+}$(4$d^4$) insulating ground state~\cite{Das2018Spin-OrbitalScattering, Arx2020}. Specifically, the excitations up to 2 eV correspond to intra-$t_{2g}$ transitions. The modes \textit{A} and \textit{B} observed at $\approx$ 80 meV and $\approx$ 350 meV, respectively, are excitations within the low-energy spin-orbital configurations with total spin $S=1$ and angular momentum $L=1$. 
The mode \textit{A} represents a final state that carries the character of a predominantly magnetic excitation, while mode \textit{B} represents a spin-orbit exciton across the lowest and highest spin-orbital sectors \cite{Das2018Spin-OrbitalScattering}. The modes \textit{C} and \textit{C}$^{\prime}$ are the manifestation of Hund's exchange (excitations from $S=1$ to $S=0$) in the channel where the final state has one doubly occupied orbital (\textit{C}) or two double occupations (\textit{C}$^\prime$), respectively 
~\cite{Arx2020,Gretarsson2019ObservationRuO_4}. The mode \textit{D} at $\approx$ 3.5 eV  is associated to $dd$ excitations, with an electron transferred across the $t_{2g}-e_{g}$ crystal field gap. 

To quantitatively describe the 
spectral weight suppression observed as a function of d.c. current bias, we introduce a fitting model using standard functional profiles that captures the above mentioned excitations, where the number of peaks and their centering are set to the $I_0$ = 0 mA case. As the d.c. current bias is increased, we fit the \gls*{RIXS} spectra allowing only the amplitude of the spectral components to replicate the observed energy-selective suppression. The result is displayed in Fig. \hyperref[fig:Fig3]{\ref*{fig:Fig3}d-f} for the largest applied d.c. current bias, while the fitting of the \gls*{RIXS} spectra measured at intermediate d.c. current values and the details of the fitting model are reported in Supplementary Fig. S2 and Section S2.2, respectively. The good agreement achieved in reproducing the \gls*{RIXS} spectra with the above fitting model corroborates the lack of peak energy shifts as a function of d.c. current. This result supports that the current-driven conductive \gls*{NESS} does not occur concomitantly with an homogeneous structural phase transition, as it is the case for the equilibrium high-$T$ metallic phase ~\cite{Braden1998CrystalTransition,Alexander1999PRB,Gorelov2010PRL,Bertinshaw2020UniqueState} and likely captured in the RIXS spectra through the $\approx$ 150 meV energy shift of the $dd$ excitation peak in Fig. \hyperref[fig:Fig2]{\ref*{fig:Fig2}c-d}.

Figure \hyperref[fig:Fig3]{\ref*{fig:Fig3}g} presents the integrated spectral weight associated with the fitted \textit{A}, \textit{B}, \textit{C}, \textit{C}$^\prime$, and \textit{D} modes as a function of d.c. current bias. The high-energy mode \textit{D}  together with the low-energy modes \textit{A} and \textit{B} show a linear suppression versus d.c. current, reaching 80-85\% of the original spectral weight for $I_4$= 2 mA. Instead, the Hund's related modes \textit{C} and \textit{C}$^{\prime}$ do not present resolvable changes. The different suppression rate experienced by the \textit{A}, \textit{B}, \textit{C}, \textit{C}$^\prime$, and \textit{D} excitations versus d.c. current is responsible for the energy-selective rescaling of the \gls*{RIXS} spectra across the current-driven \gls*{IMT}. Such peculiar energy-selective evolution of the RIXS spectral weight across the current-driven \gls*{IMT}, without any accompanying excitations energy shift or low-energy spectral weight reorganization due to the Mott gap closure, 
establishes the conductive NESS as electronically distinct from the thermal-driven metallic phase and demands for a specific non-equilibrium mechanism 
preserving the Mott gap.

\begin{figure*}
    \centering
    \includegraphics[width=1\linewidth]{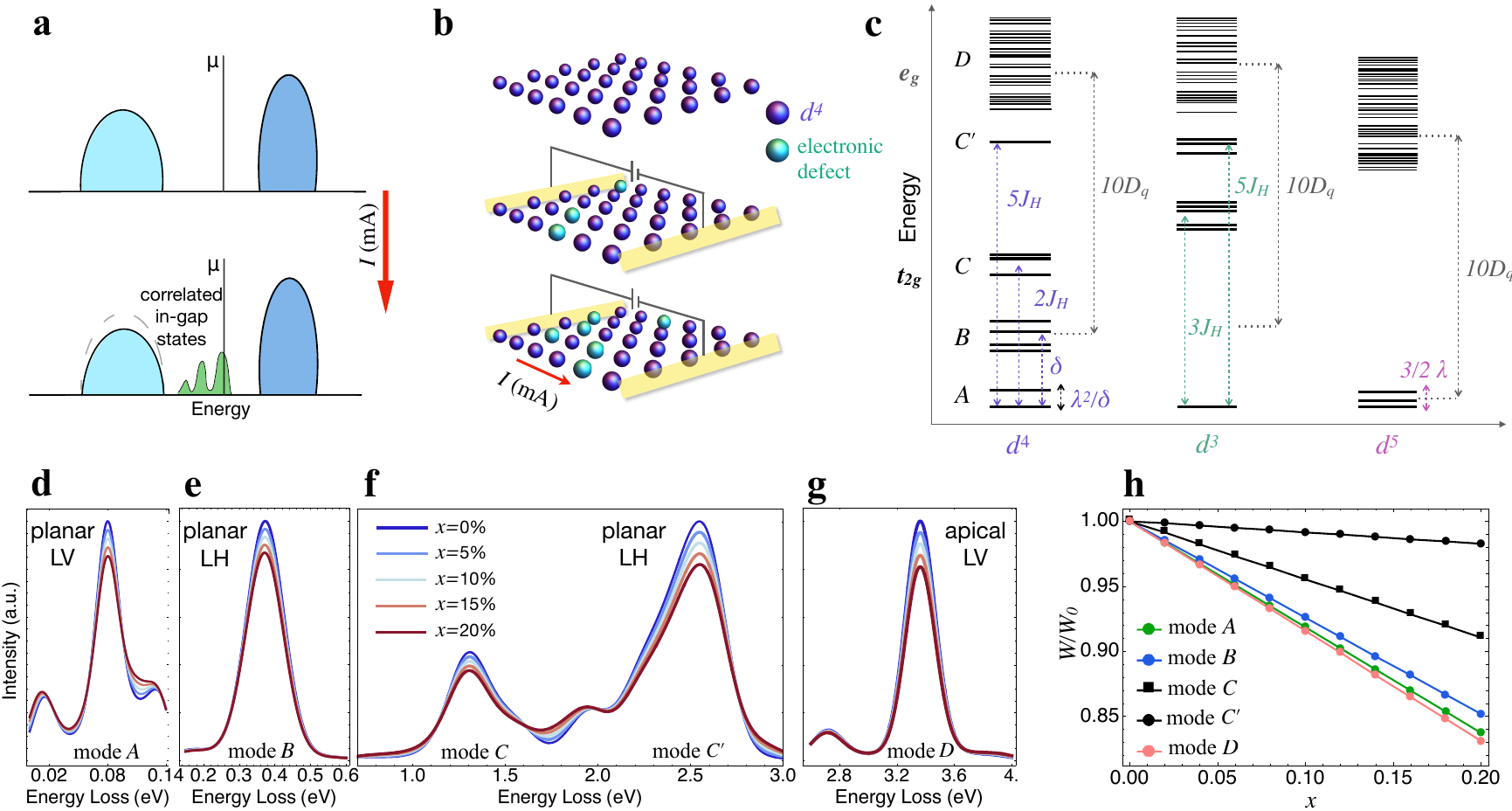}
    \caption{\textbf{Microscopic model capturing the evolution of the in-operando \gls*{RIXS} spectra across the current-driven \gls*{IMT}}. {\bf a}, Schematic representation of the formation of correlated electronic states 
    within the insulating gap. {\bf b}, Creation of electronic defect states corresponding to charge excitations in the background of the $d^4$ configurations, where 4 electrons are localized at each Ru site. Charge defects can be $d^3$, $d^5$, or $d^6$, representing one hole, one extra electron (doublon), or two extra electrons, respectively. The bottom panel illustrates that, depending on the concentration of these defects, different conductivity regimes can be achieved. {\bf c}, Multiplet structures for Ru $d^4$, $d^3$ and $d^5$ evaluated on a fully correlated model, as detailed in the Supplementary Section S3. The correspondence with the energy range of modes \textit{A}, \textit{B}, \textit{C}, \textit{C}$^{\prime}$, and \textit{D}, in which the local levels are respectively involved, is highlighted. The parameters $J_H$, 10$D_q$, $\delta$, and $\lambda$ are defined in the Methods and in Supplementary Section S3. {\bf d-g},  Theoretical calculation of the RIXS spectra as a function of of $d^3$ defects concentration $x$, at planar ({\bf d-f}) and apical ({\bf g}) resonances. The $d^3$ charge defects are considered homogeneously distributed across the sample. {\bf h}, 
    Integrated weight of the calculated RIXS spectrum $W$
    , as a function of $d^3$ defects concentration $x$. The weight of each mode is normalized to the value it has for $x=0$.} 
    \label{fig:theory}
\end{figure*}

\section{Electronic states driving insulating-to-conductive transition} \label{sec:in-gap state}

In this context, to explain the above experimental observations, 
we propose a model based on three main assumptions: i) the creation of correlated electronic states at the Fermi level within the bulk band gap of the homogeneous insulating phase, see Fig. \hyperref[fig:theory]{\ref*{fig:theory}a}; ii) such electronic states are of the type of,  e.g., 
Ru$^{5+}$($d^3$), Ru$^{3+}$($d^5$), or Ru$^{2+}$($d^6$), and 
may eventually carry the electrical current while coexisting with the Ru$^{4+}$($d^4$) insulating host state; and iii) the populations of the 
defect states are unbalanced with the $d^3$ one being the predominant. 
We sketch this situation in Fig. \hyperref[fig:theory]{\ref*{fig:theory}b}, where, the top panel shows the quasi-2D insulating ground state with only $d^4$ electronic states (blue spheres) and the central panel shows the 
charge defects (green spheres) induced by current. Upon increasing the current bias (bottom panel), the concentration of charge defects also increases, eventually enabling conductivity. 
In the following, we will show that this scenario can clarify the specific spectral weight variations observed in the RIXS spectra across the current-induced \gls*{IMT}, and we will also 
provide arguments in support of the above assumptions. 

From a spectroscopic point of view, the 
electronic defect states produce distinct multiplet structure which are characteristic of their charge configuration.
Crystal-field (CF) multiplet calculations for the configurations $d^4$, $d^3$, and $d^5$ are displayed in  Fig. \hyperref[fig:theory]{\ref*{fig:theory}c}. We do not examine the $d^6$ configuration (although it can contribute to the current flow), as it has low-energy $t_{2g}$ shell fully occupied, and thus does not contribute to the O $K$-edge RIXS process.
The resulting distribution of multiplets occurs without significantly affecting the energy separation between the $t_{2g}$ and $e_g$ sectors, denoted by $10D_q$. 
The structures of the $d^3$ and $d^5$ multiplets naturally suggest the source of variation of the RIXS spectra.
This is because the excitations of $d^3$ and $d^5$ multiplets are comparable in energy with those of $d^4$ but some modes are absent, implying an effective vanishing of the corresponding spectral weight.
Hence, in this picture $d^3$ and $d^5$ act as scattering centers with a selective variation of the RIXS spectral weight as a function of energy loss.
For instance, the lowest-energy intra-$t_{2g}$ excitations originating from the $d^3$ configuration with $S=\frac{3}{2}$ no longer include the energy sectors associated with the \textit{A} and \textit{B} modes of the $d^4$ host. Instead, these excitations now involve high-to-low spin transitions that occur at higher energies, dictated by the Hund's coupling parameter $3J_{H}$ (further details in Supplementary Section S3).\\

While the multiplet description already provides a clear-cut qualitative picture of the energy selective variation of the \gls*{RIXS} spectra,  
we numerically calculate the O $K$-edge RIXS spectrum of a cluster including $d^3$, $d^4$ and $d^5$ configurations to quantitatively account for the coupling between the defects and the $d^4$ host. We consider a multi-orbital microscopic model which is the commonly adopted for correlated electrons in Ca$_{2}$RuO$_4$ (See Supplementary Section S3). We compute the RIXS cross-section using exact diagonalization calculations (ED) within the fast collision approximation (See Methods Section \ref{sec:TheoreticalModeling} and Supplementary Section S3.3), for different clusters of Ru ions bonded via planar or apical oxygens. The simulated RIXS intensity is derived as the weighted average of the independent contributions from both defect-free (insulating ground state at 0 mA) and defect-containing clusters (conductive \gls*{NESS}  state). Here, we consider that the magnitude of the applied current scales linearly with the defect concentration $x$. In the calculation shown in Fig. \hyperref[fig:theory]{\ref*{fig:theory}d-g}, we assume that the concentration $x$ is exclusively due to $d^3$ defects. This choice is based on the initial assumption of a predominant $d^3$ concentration. 
Therefore, 
we express the RIXS cross-section as $\mathcal{I}_{total}(x)=(1-x) \mathcal{I}_{d^4}+x \mathcal{I}_{d^3}$. Defect-free clusters  produce the characteristic spectral features ($\mathcal{I}_{d^4}$) of the Ca$_{2}$RuO$_4$ insulating phase and the clusters with defects introduce alterations in the electronic structure with distinct spectral features ($\mathcal{I}_{d^3}$). By combining the $d^4$ ground state and the $d^3$ charge defects contribution, the simulation 
reasonably 
captures the impact of charge defects on the RIXS spectrum.  Figures \hyperref[fig:theory]{\ref*{fig:theory}d-g} show the theoretically simulated RIXS cross-section at $T$ = 0 K, analyzed as a function of $d^3$ defect concentration $x$ within the material. Each curve represents the RIXS spectrum for a different impurity concentration. For increasing impurity concentration $x$,  a reduction in the  calculated primary RIXS excitations intensity is observed. This reduction predominantly affects excitations $A$, $B$, and $D$, and it can be attributed to the scattering centers introduced by the $d^3$ defects which inhibit electronic transitions and reduce the overall spectral weight of those excitations. In contrast, for modes \textit{C} and \textit{C}$^\prime$ reflecting the Hund's exchange, the impurity-induced scattering creates a more complex landscape of electronic transitions, leading to a distribution of energy levels across a wider range of energy losses effectively resulting in a broadening. 

To compare our key experimental findings -- energy selective 
suppression of the RIXS spectral weight -- with the proposed microscopic mechanism, we compute the variation of the spectral weight $W$ of each mode as a function of $x$. The integral of the RIXS intensity over the energy range (${E_1}-{E_2}$) of a mode $j$ can be written as $W_{j }(x)= \int_{E_1}^{E_2} \mathcal{I}_{total}(x,E) \,dE/\int_{E_1}^{E_2} \mathcal{I}_{d^4}(E) \,dE $, where $E$ is the energy loss (details reported in Supplementary Section S3.3).
The result (Fig. \hyperref[fig:theory]{\ref*{fig:theory}h}) suggests that the spectral weights vary linearly with $x$, and exhibits a marked decrease for the modes  \textit{A}, \textit{B}, and \textit{D}, as observed experimentally in Fig. \hyperref[fig:Fig2]{\ref*{fig:Fig2}g}. At variance with the experiment, however, the calculations present a small spectral weight suppression for the features \textit{C} and \textit{C}$^\prime$, not present in the \gls*{RIXS} spectra. 
We attribute this discrepancy to the fact that the calculations are performed at $T= 0$ K, while the experiment is conducted at room temperature, where the features \textit{C} and \textit{C}$^\prime$ are heavily broadened with respect to the low-temperature data \cite{Das2018Spin-OrbitalScattering} due to their sensitivity to thermal effects, thus diluting the effect of the ${d^3}$ defects. 
Overall, the calculated trend in Fig. \hyperref[fig:theory]{\ref*{fig:theory}h} is in good qualitative agreement with the experimental data in Fig. \hyperref[fig:Fig3]{\ref*{fig:Fig3}g}, validating the assumptions that the $d^3$ concentration is predominant and linearly increases with the current passing through the material, and endorsing 
the electronic origin of the conductive \gls*{NESS}. We hence propose that the correlated electronic defect states 
act as a potential driving mechanism of the current-induced \gls*{IMT} in Ca$_{2}$RuO$_4$. 

We finally present a physical scenario that can account for the hypothesis of having current-induced correlated electronic states. 
As explained in detail in the Supplementary Section S3.2, a combination of octahedral distortions and modifications to short-range Coulomb interactions can lead to the emergence of 
$d^3$, $d^5$, and $d^6$ correlated states within the Mott gap. These defects, situated at the Fermi level of the host insulating band structure, facilitate the onset of conductivity, without collapsing the Mott gap. Given the differing phase space for electronic configurations (see Supplementary Fig. S4), the population of $d^3$ states can exceed that of $d^5$ and $d^6$ states, supporting the modeling hypothesis for the evolution of the RIXS spectra.
We propose that the enhancement of the short-range Coulomb interactions may arise at the interface between short and elongated octahedra, either naturally present in the crystal or induced by current \cite{Gauquelin2023NanoLett}. 
This can be also accounted by the fact that the breaking of inversion symmetry in these regions can lead to polar effects, which are known to induce an extra term of nearest neighbor Coulomb interaction \cite{meinders1995atomic,van1995influence}. 
\section{Conclusions}

We performed in-operando \gls*{RIXS} investigations of Ca$_2$RuO$_4$ across the current-driven \gls*{IMT}, and observed an energy selective suppression of the spectral weight as a function of d.c. current, without involving shifts in the energy spectrum or reorganization of the low-energy spectral weight. By combining our experimental results with RIXS cross-section calculations, we demonstrated that the d.c. current applied across the material creates electronic defects at the Ru sites, 
and that the concentration of such defects increases with the supplied d.c. current. We thus conclude that the current-driven conductive non-equilibrium steady state has an electronic origin, and can persist at the presence of a Mott insulating band gap. Complementary RIXS measurements of the thermal-driven \gls*{IMT} further support the need of a distinctive mechanism behind the current-driven \gls*{IMT}, ruling out at the same time Joule-heating effects induced by the d.c. current bias.

The framework presented in this study can be applied to probe electrically-induced phase transition or manipulation in other strongly correlated materials. The in-operando \gls*{RIXS} technique provides a powerful approach to investigate at a microscopic and bulk level the interplay between the electronic degrees of freedom 
across electrically-induced phase transitions. 
Moreover, understanding the mechanisms driving \gls*{IMT} in strongly correlated systems not only deepens our fundamental knowledge, but also has practical implications for the design of materials in areas such as spintronics and orbitronics. These applications demand precise control over electronic states, making insights into the underlying physics particularly valuable.

\acknowledgments
This work was supported by the US Department of Energy (DOE) Office of Science, Early Career Research Program. The in-operando \gls*{RIXS} setup was co-supported as part of Programmable Quantum Materials, an Energy Frontier Research Center funded by the U.S. Department of Energy (DOE), Office of Science, Basic Energy Sciences (BES), under award DE-SC0019443. 
This research used beamline 2-ID of NSLS-II, a US DOE Office of Science User Facility operated for the DOE Office of Science by Brookhaven National Laboratory under contract no. DE-SC0012704. This research used the Electron Microscopy facility of the Center for Functional Nanomaterials (CFN), which is a U.S. Department of Energy Office of Science User Facility, at Brookhaven National Laboratory under Contract No. DE-SC0012704. S.F. and J.P. were supported by the Laboratory Directed Research and Development project of Brookhaven National Laboratory No. 21-037. M.C., R.F., M.L., A.D.B., and A.V. acknowledge support from the EU’s Horizon 2020 research and innovation program under Grant Agreement No. 964398 (SUPERGATE). M.C., F.F. and G.C. acknowledge support from PNRR MUR project PE0000023-NQSTI. A.D.B. also acknowledges support from the MAECI project 'ULTRAQMAT' and, together with R. H., support from the Alexander von Humboldt Foundation in the framework of a Sofja Kovalevskaja grant. R.H. acknowledges support from the DFG under project no. 493158779 as part of the collaborative research project SFB F 86 Q-M\&S funded by the Austrian Science Fund (FWF) project number LAP 8610-N. This research has been supported by PNRR MUR project PE0000023-NQSTI. This work was supported in part by the Italian Ministry of Foreign Affairs and International Cooperation, grant number KR23GR06. We acknowledge the use of the experimental equipment and the expert support concerning its usage provided by the Nanostructure Laboratory at the University of Konstanz.
F.F. and F.G. acknowledge support from the Italian Ministry of University and Research (MUR) under Grant PRIN No. 2020JZ5N9M (CONQUEST) and under the Grant PRIN 2022, funded by the European Union - NextGenerationEU, Mission 4, Component 2, Grant No. 2022TWZ9NR (STIMO)-CUP B53D23004560006. C.A. was supported by the Foundation for Polish Science project “MagTop” no. FENG.02.01-IP.05-0028/23 co-financed by the European Union from the funds of Priority 2 of the European Funds for a Smart Economy Program 2021–2027 (FENG). We acknowledge the access to the computing facilities of the Interdisciplinary Center of Modeling at the University of Warsaw, Grant g91-1418, g91-1419, g96-1808 and g96-1809 for the availability of high-performance computing resources and support. We acknowledge the CINECA award under the ISCRA initiative IsC99 ``SILENTS”, IsC105 ``SILENTSG" and IsB26 ``SHINY" grants for the availability of high-performance computing resources and support. We acknowledge the access to the computing facilities of the Poznan Supercomputing and Networking Center Grant No. 609.\\

\section{Author Contributions}
VB, FF, MC, AV, ADB, CM conceived the research project. VKB, VB, RH, ADB, AV, and MC designed the study, with contributions from RF, CM, and JP. AV, RF, and ML grew the CRO single crystals. RH fabricated and characterized the CRO devices with help from ADB, VKB, FC, KK, DNB, AV, and VB. VKB, TK, SF, JP, RH, ADB, and VB carried out the RIXS measurements and performed the first data interpretation. VKB analyzed the RIXS data with guidance from VB. FF, FG, and MC contributed to the data discussion. FF performed the RIXS cross-section calculations to interpret the experimental observations, FG analyzed the theoretical model of charge defects, with the help of MC. GC and CA performed the ab-initio calculation. VKB, RH, FF, MC, and VB wrote the manuscript with contributions from all authors.

\section{Competing interests}

The authors declare no competing interests.

\section{Additional Information}
{\bf{Supplementary Information}} is available for this paper at ...

{\bf{Correspondence and requests for materials}} should be addressed to V. Bhartiya, F. Forte, or V. Bisogni.

\bibliography{VB_Library}

\clearpage
\section{Methods} 
\subsection{Sample Information}\label{sec:device_preparation}
For the devices, we used flakes of Ca$_2$RuO$_4$ obtained by exfoliating high-quality Ca$_{2}$RuO$_4$ single crystals with the scotch-tape technique. The Ca$_{2}$RuO$_4$ single crystals were grown by the floating zone technique and their structural and electronic properties had been thoroughly investigated in previous studies using a variety of techniques including RIXS \cite{Das2018Spin-OrbitalScattering,Curcio2023Current-drivenCa2RuO4,Fatuzzo2015Spin-orbit-inducedStudy}. 

\subsection{X-ray Absorption Spectroscopy and Resonant Inelastic X-ray Scattering} \label{sec:spectroscopy}
 The \gls*{XAS} and high-resolution O $K$-edge in-operando \gls*{RIXS} experiments were performed at the SIX 2-ID beamline of NSLS-II using the Centurion RIXS spectrometer~\cite{DovrakRSI2016}. The combined energy resolution at the O $K-$ edge ($\approx$ 530 eV) at apical resonance was $\Delta E$  = 25 meV  (full-width at half-maximum, measured at a multi-layer reference sample). For the planar resonance, the lowest energy mode is the spin-orbit exciton centered at $\approx$ 350 meV, therefore, to optimize the measurement time, we compromised on resolution and, a twice larger FWHM  was employed. Throughout the experiment, both $\pi$ and $\sigma$-polarized X-rays were used.  The temperature was fixed to 300 K for both the \gls*{XAS} and RIXS measurements. As shown in the main text of Fig.\hyperref[fig:Fig1]{\ref*{fig:Fig1}b}, the scattering geometry was kept fixed throughout the experiment, with the incident angle $\theta$ = 45$^\circ$ and the scattering angle $2\theta$ = 150$^\circ$. For the device, the sample out-of-plane $c$-axis lied in the scattering plane. The two planar axes $a$ and $b$ could not be distinguished from each other, hence, represented as $a/b$. We found that the $a/b$-axis makes an angle of $\sim$ 34$^\circ$ with respect to the horizontal scattering plane.  
 For the single crystal, the sample out-of-plane $c$-axis and in-plane $a$-axis lied in the scattering plane.

\subsection{Theoretical Modeling} \label{sec:TheoreticalModeling}
RIXS spectra were generated by the exact diagonalization of two different types of finite-sized clusters. The first is composed of two Ru atoms, each bonded to a central planar oxygen (Ru$_1$-O$^p$-Ru$_2$), and the second is composed of a Ru site connected to an apical oxygen (Ru-O$^{ap}$). We employ a microscopic Hamiltonian that describes both the O $(p_x,p_y,p_z$) and Ru $(d_{yz}, d_{zx}, d_{xy}, d_{x^2-y^2}, d_{z^2}$) orbitals.  This Hamiltonian is projected onto the subspace of the $t_{2g}$ orbitals in the case of the first cluster, and onto both the  $t_{2g}$ and $e_{g}$ manifold in the case of the second cluster. The model incorporates several key components: the on-site Coulomb interactions among $d$ electrons, and the hybridization between $d$ and the planar or apical oxygen $p$ bands. Additionally, it includes the CF splitting, which affects the energy levels of the $d$ orbitals, and the spin-orbit coupling (SOC), which mixes spin and orbital degrees of freedom. 
The values of the energies of the 4$d$ orbitals, specifically the hopping and crystal field parameters (the average $e_g$ to $t_{2g}$ splitting 10$D_q$, the $t_{2g}$ tetragonal distortion $\delta$, and the orthorhombic distortion $\delta_{ort}$), were estimated with the support of ab initio calculations, as detailed in the Supplementary Section S3.5. The other microscopic parameters used to simulate the RIXS spectra are aligned with values found in prior studies \cite{Das2018Spin-OrbitalScattering,Arx2020}. Specifically, the on-site Coulomb interaction $U$ is typically set to approximately 2$-$2.2 eV, while the Hund's coupling $J_H$ is around 0.4 eV. For the local spin-orbit coupling, a value of $\lambda \sim $ 0.07 eV has been employed in multiple works~\cite{Jain2017HiggsAntiferromagnet,Gretarsson2019ObservationRuO_4, Fatuzzo2015Spin-orbit-inducedStudy}. Additionally, the system assumes an orthorhombic unit cell configuration where in-plane Ru-O-Ru bond angles deviate from 180$\degree$, due to rotations of the RuO$_6$ octahedra, both around the $c$ and in-plane axis. These bond angles fall within a range of approximately 170\degree, due to these distortions ~\cite{Porter2018MagneticRuO_4}. More details are presented in the Supplementary Section S3.
\end{document}




\title{\textbf{Supplementary Information for ``Evidence of electronic states driving current-induced insulator-to-metal transition'' }}

\author{V. K. Bhartiya}
\email{vbhartiya1@bnl.gov}
\affiliation{National Synchrotron Light Source II, Brookhaven National Laboratory, Upton, New York 11973, USA}
\author{R. Hartmann}
\affiliation{Department of Physics, University of Konstanz, 78457 Konstanz, Germany}
\author{F. Forte}
\email{filomena.forte@spin.cnr.it}
\affiliation{CNR-SPIN, c/o Universit\`a di Salerno, I-84084 Fisciano, Salerno, Italy}
\author{F. Gabriele}
\affiliation{CNR-SPIN, c/o Universit\`a di Salerno, I-84084 Fisciano, Salerno, Italy}
\author{T. Kim}
\affiliation{National Synchrotron Light Source II, Brookhaven National Laboratory, Upton, New York 11973, USA}
\author{G. Cuono}
\affiliation{CNR-SPIN, c/o Universit\`a “G. d\textquotesingle Annunzio”, I-66100 Chieti, Italy }
\author{C. Autieri}
\affiliation{International Research Centre Magtop, Institute of Physics, Polish Academy of Sciences, Aleja Lotnik\'ow 32/46, 02668 Warsaw, Poland}
\author{S. Fan}
\affiliation{National Synchrotron Light Source II, Brookhaven National Laboratory, Upton, New York 11973, USA}

\author{K. Kisslinger}
\affiliation{Center for Functional Nanomaterials, Brookhaven National Laboratory, Upton, New York 11973, USA }

\author{F. Camino}
\affiliation{Center for Functional Nanomaterials, Brookhaven National Laboratory, Upton, New York 11973, USA }

\author{M. Lettieri}
\affiliation{CNR-SPIN, c/o Universit\`a di Salerno, I-84084 Fisciano, Salerno, Italy}

\author{R. Fittipaldi}
\affiliation{CNR-SPIN, c/o Universit\`a di Salerno, I-84084 Fisciano, Salerno, Italy}

\author{C. Mazzoli}
\affiliation{National Synchrotron Light Source II, Brookhaven National Laboratory, Upton, New York 11973, USA}

\author{D. N. Basov}
\affiliation{Department of Physics, Columbia University, New York, New York 10027, USA.}

\author{J. Pelliciari}
\affiliation{National Synchrotron Light Source II, Brookhaven National Laboratory, Upton, New York 11973, USA}

\author{A. Di Bernardo}
\affiliation{Department of Physics, University of Konstanz, 78457 Konstanz, Germany}
\affiliation{Universit\`a di Salerno, Department of Physics ``E. R. Caianiello", 84084 Fisciano, Salerno, Italy}

\author{A. Vecchione}
\affiliation{CNR-SPIN, c/o Universit\`a di Salerno, I-84084 Fisciano, Salerno, Italy}
\author{M. Cuoco}
\affiliation{CNR-SPIN, c/o Universit\`a di Salerno, I-84084 Fisciano, Salerno, Italy}
\author{V. Bisogni}
\email{bisogni@bnl.gov}
\affiliation{National Synchrotron Light Source II, Brookhaven National Laboratory, Upton, New York 11973, USA}

\maketitle
\tableofcontents
\clearpage
\section{Device}\label{sec:device}
\subsection{Fabrication and characterization} \label{sec:device_preparation}
The devices were fabricated starting from Ca$_{2}$RuO$_4$ flakes exfoliated from high-quality single crystals, using the scotch-tape technique, and deposited on a SiO$_2$ (300 nm)/Si substrate with pre-patterned indexed Au markers (to facilitate the identification of the flakes and their localization in the following fabrication steps). The Ca$_{2}$RuO$_4$ flakes were mapped under an optical microscope to select the ones with a thickness of few micrometers and lateral size of $\approx$  10  $\times$ 10 \SI{}{\micro\meter}$^2$. 
After selecting flakes of the desired size, electrical contacts were deposited to realize a two-point configuration.  
Electrical contacts to the flakes were made by defining their geometry using electron beam lithography (Zeiss Crossbeam 1540XB SEM in combination with Neomicra Smile 2 lithography system) on a spin-coated double-layer polymethyl methacrylate A4/methyl methacrylate EL11 (PMMA A4/MMA-MAA EL 11) resist. As shown in Fig. 1{\bf a} of the main text, contact electrodes ($\approx$ 12 \SI{}{\micro\metre}- wider than the flakes) were designed to maximize the fraction of sample volume crossed by the bias current and to ensure good electrical contact. These electrodes were made by sputtering on a freshly-cleaved Ca$_{2}$RuO$_4$ surface after a soft Ar milling performed in-situ in the sputtering chamber. Sputtering was chosen as the deposition technique over thermal evaporation to ensure a better coverage of the step at the edge of the flakes compared to evaporation. The contact leads consist of $\approx$ 250 nm of Ti followed by $\approx$ 50 nm of Au. After lift off and removal of the residual in acetone, the lateral edges of the Au contacts were also reinforced with Pt conducting patches; which were deposited via focused ion-beam induced deposition (FIBID) with a Gallium beam. The deposition was carried out with an acceleration voltage of 30 kV and current 50 pA in a dual-beam focused ion beam/scanning electron microscope (FIB/SEM, Thermo Scientific Helios G5 and Zeiss Crossbeam 1540XB). Extreme care was taken to avoid exposure (and related damage) of the inter-electrode area of the sample surface. After deposition, a gentle low-current ion-beam milling (acceleration voltage of 2 kV, current of 50 pA) of the sample surface was performed to remove possible Pt oversprayed in the inter-electrode area during the Pt deposition.
The thickness  of the Ca$_{2}$RuO$_4$ flake ($\approx$ 1 \SI{}{\micro\metre} of device used in this investigation) was determined using a Bruker Multimode atomic force microscope (AFM) in tapping mode and Tap300Al-G probes by BudgetSensors.

Initial electrical characterization was performed either by bonding the device with Al wires to a chip holder or by contacting it in a MPI ITS50-COAX wafer prober with tungsten tips. Measurements were done with a Keithley 6221 current source in combination with a Keithley 2182A nano-voltmeter in the case of bonded samples or with a Keithley 2450 source meter in combination with the wafer prober. In both cases a current bias was applied to the device and the voltage drop was measured across the flake in a two-point configuration.
 
\subsection{Flake crystallographic orientation}\label{sec:device_orientation}
The scanning electron microscope (SEM) micrographs of the Ca$_{2}$RuO$_4$ device were collected using a Zeiss Gemini 500 SEM with an acceleration voltage  of 10 kV, see Fig. 1a of the main text. The in-plane crystallographic orientation of the flakes was determined through the same SEM setup, using the electron back-scattering diffraction (EBSD) with an  Oxford Instrument SYMMETRY detector and the Oxford Aztec software for analysis of the detected diffraction patterns at 20 kV acceleration voltage.
Ca$_{2}$RuO$_4$ cleaves along the [001] plane, such that the flake surface is perpendicular to the \textit{c}-axis. The SEM measurements could identify the orientation of the in-plane axes \textit{a}/\textit{b}, but could not discriminate between them due to the small tetragonal distortion. 

\section{In-operando \gls*{RIXS}} \label{sec:oprixs}
\subsection{ I-V measurements}\label{sec:IV_charactristics}

\begin{figure}
\centering
\includegraphics[width=0.9\linewidth]{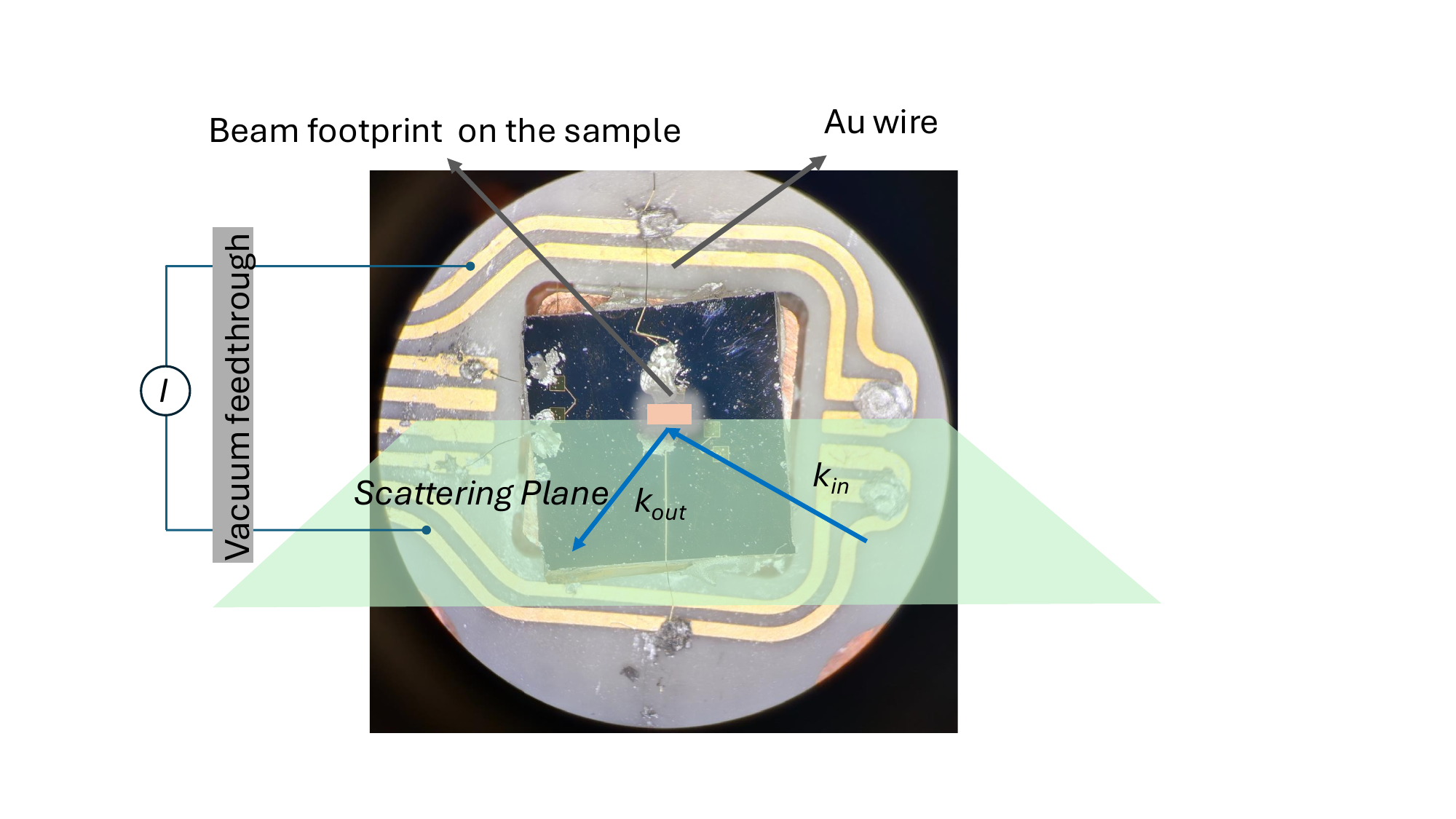}
\caption{The in-operando \gls*{RIXS} set up of the 2-ID beamlime at NSLSII. A device of $\approx$ 10 $\times$ 10 $\times$ 1 \SI{}{\micro\metre}$^3$ size is exposed with $\approx$ 10 $\times$ 2 \SI{}{\micro\metre}$^2$ beamsize while driving d.c. current through the device.}
\label{fig:operando_device}
\end{figure}

 The in-operando room temperature I-V characteristic was measured in a two-wire configuration with the Keithely2636B source meter (Fig. \ref{fig:operando_device}). Device was electrically connected through Au wire (diameter 17 \SI{}{\micro\metre}). One end was glued to the gold electrodes with the silver paint and another end to a special custom made sample holder connected to the source meter through special feed through cables connecting vacuum and the source meter. The voltage drop was recordedas the current was changed during the in-operando \gls*{RIXS} measurements. 

\subsection{RIXS spectrum fitting} \label{sec:RIXS_spectrum_modelling}
We accounted for the O $K$-edge \gls*{RIXS} spectrum from $-$2 eV to 7 eV energy transfer range in the modeling. A representative \gls*{RIXS} scan (at planar resonance) is shown in Fig. \ref{fig:Fig_reversal}. The lowest mode of interest (mode $A$) is resonant only at the apical resonance; therefore, high-resolution in-operando \gls*{RIXS} (FWHM $\approx$ 24 meV) was collected only for the apical oxygen resonance.  For the apical resonance \gls*{RIXS} spectrum, elastic line was simulated with a Psuedo-Voigt lineshape and mode $A$  was simulated with a Gaussian lineshape. Moreover, two more low-energy modes centered at 25 meV and 40 meV could also be resolved and were also modeled (not shown in the figures) with a resolution limited Gaussian. For the planar oxygen resonance, a broader lineshape (FWHM $\approx$ 50 meV) was used to account for the elastic lineshape.  A multi-component spin-orbit exciton  was simulated with a low-energy (120 meV - not shown in the figures) Gaussian and the high-energy Skewed-Gaussian. The broad profile around 3 $-$ 7 eV is made of two $dd$- excitations (one resonating at apical resonance and another resonating at planar resonance) and possible Ca-O bands contribution, therefore to account for this high energy spectral weight, three Skewed-Gaussian lineshapes were used. Peak positions of all the modes were fixed to the values to the ground state at 0 mA. The exact expressions of the used lineshapes are as follows:

Psuedo-Voigt profile for the elastic line:

\begin{equation}
\begin{split}
    f_1(x;A,\mu,\sigma,\alpha) &= \frac{(1-\alpha)A}{\sigma_g \sqrt{2\pi}}e^{-(x-\mu)^2/2\sigma_g^2} \\&+ \frac{\alpha A}{\pi} \frac{\sigma}{(x-\mu)^2+\sigma^2}
    \end{split}
\end{equation}

Gaussian lineshape for low energy phonon and spin-orbital magnetic excitation:

\begin{equation}
    f_2(x;A,\mu,\sigma) = \frac{A}{\sigma \sqrt{2 \pi}} e^{[-(x-\mu)^2/2\sigma^2]}
\end{equation}

\begin{figure*}
    \centering
    \includegraphics[width=1\linewidth]{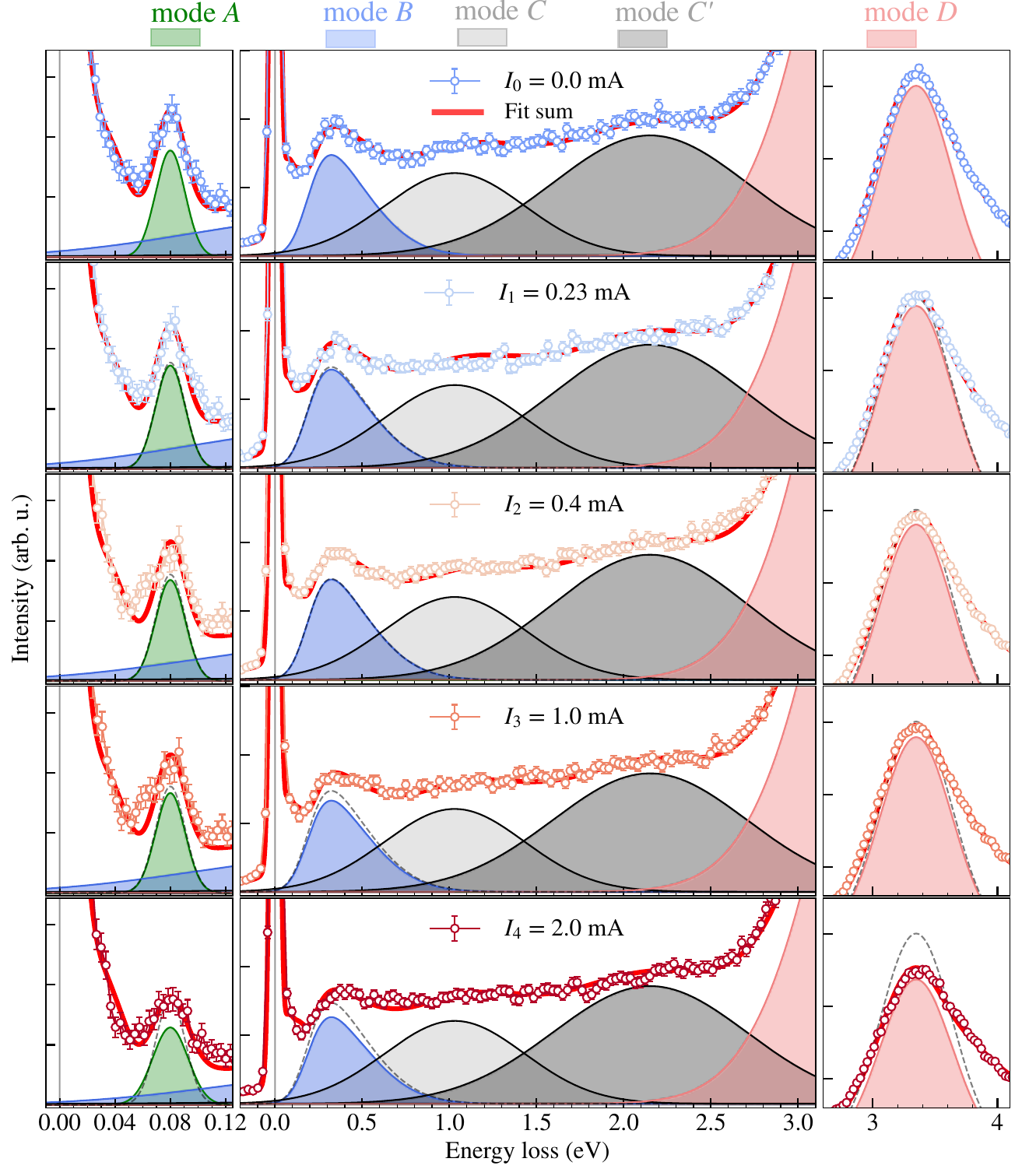}
    \caption{\textbf{A complete set of \gls*{RIXS} spectrum across current-driven \gls*{IMT} at 300 K}. Left and right most plots in an horizontal panel (at a given  current $I$) are measured at the apical resonance and corresponding central panel at the planar resonance. All the fitted modes shown here are defined in Fig. 3 and section 2 of the main text.}
    \label{fig:Idepsummary}
\end{figure*}

\begin{figure}
\centering
\includegraphics[width=1\linewidth]{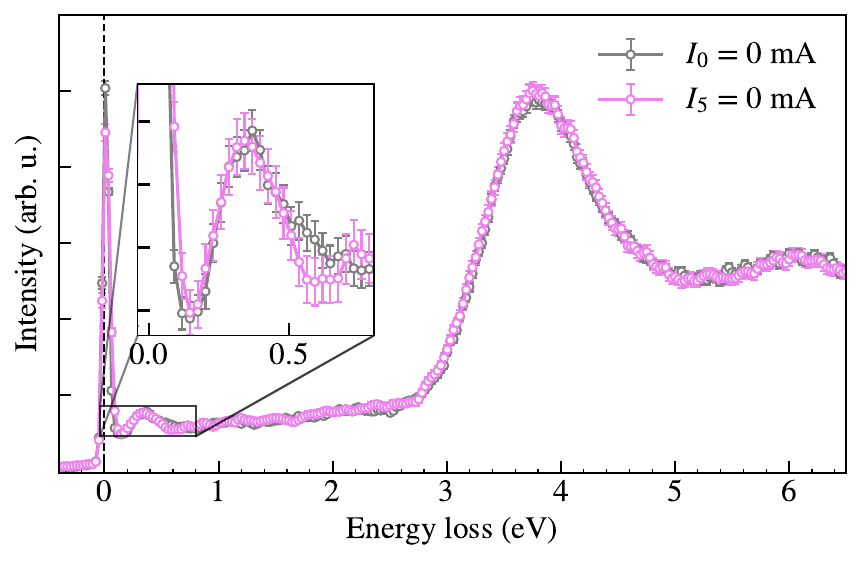}
\caption{\textbf{Complete recovery of \gls*{RIXS} spectrum upon returning to $I_5 = $ 0 mA}. The gray colored \gls*{RIXS} spectrum was measured at  $I_0 = $ 0 mA before activating the device and the violet colored \gls*{RIXS} spectrum was measured at $I_5$ after deactivating (by quickly ramping down the current from $I_4 = $ 2 mA to $I_5 = 0$ mA). Inset highlights a complete recovery of spin-orbit exciton. Full recovery to the insulting ground state after a week long operation show that the device did not incur permanent changes/damage and the current induced conductive state goes away with current-- a direct microscopic signature of \gls*{NESS}~\cite{Nakamura2013Electricfield,Bertinshaw2020UniqueState}.}
\label{fig:Fig_reversal}
\end{figure}

Skewed Gaussian for spin-orbit exciton and $dd$-excitations:
\begin{equation}
    f_3(x;A,\mu,\sigma,\gamma) = \frac{A}{\sigma \sqrt{2 \pi}} e^{[-(x-\mu)^2/2\sigma^2]} \{1+\text{erf}[\frac{\gamma(x-\mu)}{\sigma\sqrt{2}}] \}
\end{equation}
Here $A$ is the amplitude, $\mu$ is the centre, $\sigma$ is the standard-deviation, $\alpha$ is the fraction, and $\gamma$ is the Skewness. The center $\mu$ was fixed for all modes as the scattering geometry (momentum transfer) was fixed during the measurements. $\sigma$ was also fixed except mode $A$ at 2 mA, as a suppressed and broad peak required a $\sim$ 20 $\%$ larger FWHM to fully account the spectral weight. A complete set of fitted \gls*{RIXS} spectra are shown in Fig. \ref{fig:Idepsummary}.

\clearpage


\clearpage

\section{Theoretical Modeling} \label{sec:theory_model}

\subsection{Model Hamiltonian} 

We adopted a multi-orbital Hubbard model to describe the Ru and O bands near the Fermi level, incorporating hopping terms connecting $d$ and $p$ orbitals of nearest-neighbor Ru and O ions, and crystal field (CF) splittings. The model also includes the interaction terms at the Ru site, and specifically the Coulomb interactions and atomic spin-orbit coupling (SOC) term. The clusters considered are of the Ru-O$^{pl}$-Ru type, bonded along the 
$x$-axis for planar oxygen, and of the Ru-O$^{ap}$ type for bonds with an apical oxygen along the $z$-direction. In the first case, the $d$-orbital manifold corresponds to the $t_{2g}$ subspace, while in the second case, it includes both $t_{2g}$ and $e_g$. The Hamiltonian has been applied to model the case in which the Ru $d$ bands have a fixed electron occupancy of $n=4$, corresponding to the system at equilibrium. Additionally, the model was solved considering a non-homogeneous filling of the Ru bands, corresponding to a configuration where charge defects ($d^3$ or $d^5$) are introduced upon gating, due to the addition or removal of charge carriers, as outlined in the hypothesis detailed in the main text.\\
The local Hamiltonian at each Ru site is defined as $H^{Ru}_{\mathrm{loc}}(i)$ and includes both the intra-orbital and inter-orbital Coulomb interaction, the SOC and CF potential \cite{Oles05,Cuoco06a,Cuoco06b}:

\begin{equation}
H^{Ru}_{\mathrm{loc}}(i)=H_{e-e}(i)+H_{\mathrm{SOC}}(i)+H_{\mathrm{CF}}(i)\, .
\label{eq:Hloc}
\end{equation}

Here, the on-site Coulomb, SOC and CF terms are expressed at
site $i$ by
\begin{eqnarray}
H_{e-e}(i)&=& U \sum_{i,\alpha} n_{i \alpha \uparrow} n_{i  \alpha \downarrow} \\ \nonumber
&+ &\sum_{i, \alpha \neq \beta , \sigma \neq \sigma'} \left( U_{\alpha \beta} - \frac{1}{2}J_{\alpha \beta} \right) n_{i \sigma} n_{i \sigma'} \\ \nonumber
&+& \sum_{i, \alpha \neq \beta, \sigma \neq \sigma'} J_{\alpha \beta} \left( d_{i \alpha \uparrow}^\dagger d_{i \alpha \downarrow}^\dagger d_{i \beta \downarrow} d_{i \beta \uparrow} + h.c.\right) \\ \nonumber 
&-& 2 \sum_{i, \alpha \neq \beta,\sigma \neq \sigma'} J_{\alpha \beta} \mathbf{S}_{i \alpha} \cdot \mathbf{S}_{i\beta} \label{eq:HCoul}
\\
H_{\mathrm{SOC}}(i)&=&\lambda \sum\limits_{\alpha ,\sigma }
\sum_{\beta ,\sigma^{^{\prime }}}\,                                                                                    d_{i\alpha \sigma }^{\dagger }\,({\overrightarrow{l}})_{\alpha\beta}
\cdot ({\overrightarrow{s}})_{\sigma \sigma ^{^{\prime }}}\,
d_{i\beta\sigma ^{^{\prime }}},  \\
H_{\mathrm{CF}}(i) &=&\varepsilon_{xy} n_{i,xy}+\left(\varepsilon_{xz}
n_{i,xz}+\varepsilon_{yz} n_{i,yz}\right) \nonumber \\
&+&\varepsilon_{x^2-y^2} n_{i,x^2-y^2}+\varepsilon_{z^2} n_{i, z^2}\, .
\end{eqnarray}
\\
%
The term $d_{i\alpha\sigma}^{\dagger}$ is the creation operator for an electron with spin $\sigma$ at site $i$ in orbital $\alpha$. The interaction parameters include the intra-orbital Coulomb interaction $U$ and
exchange elements, $U_{\alpha \beta}$ and $J_{\alpha \beta}$, which in general are anisotropic and depend
on the involved pair of orbitals. The Coulomb and exchange elements are connected to the intra-orbital element $U$ by a relationship $U=U_{\alpha \beta}+2 J_{\alpha \beta}$, that ensures the invariance of interactions within the orbital space. In scenarios where only one type of orbital (either $e_g$ or $t_{2g}$) is partially occupied, all relevant exchange elements $J_{\alpha,\beta}$ are identical. In such cases, a simplified form of onsite interactions can be used using only two parameters: the Coulomb element $U$ and a Hund’s exchange element $J_H$. Otherwise, these parameters will be orbital dependent. \cite{Oles05}
The CF term $H_{\mathrm{CF}}(i)$ describes the extent of the splitting between the energies $\epsilon_{d}$ of the $d$ levels, and can be expressed in terms of two main parameters: $10D_{q}$, which quantifies the separation between the average energy of the $e_{g}$ and that of the $t_{2g}$ orbitals; $\delta = (\varepsilon_{xy} - \varepsilon_z)$ which parametrizes the strength of the tetragonal distortions within the $t_{2g}$ (the same splitting is assumed within the $e_{g}$ sector). We also consider orthorhombic splitting, $\delta_{ort}$, of the ${xz, yz}$ orbitals, assuming $\varepsilon_{yz} = \varepsilon_z + \delta_{ort}$ and $\varepsilon_{zx} = \varepsilon_z - \delta_{ort}$.\\

Due to the large value of $10D_{q}$ ($\sim$3 eV), the spin-orbit term is effectively described by projecting out the $e_g$ levels, which are well above
the t$_{2g}$ levels. The operator $\overrightarrow{l}$ restricts the angular momentum operators to the $t_{2g}$ subspace, and $\overrightarrow{s}_i = \frac{1}{2} \overrightarrow{\sigma}_i$ is the spin operator at site $i$ -- expressed through the Pauli matrices $\overrightarrow{\sigma}_i$.

\begin{eqnarray*}
l_{x} &=&
\begin{bmatrix}
0 & 0 & 0 \\
0 & 0 & i \\
0 & -i & 0%
\end{bmatrix}
\rightarrow
\begin{bmatrix}
d_{yz} \\
d_{xz} \\
d_{xy}%
\end{bmatrix}%
\  \\
\ l_{y} &=&%
\begin{bmatrix}
0 & 0 & -i \\
0 & 0 & 0 \\
i & 0 & 0%
\end{bmatrix}%
\rightarrow
\begin{bmatrix}
d_{yz} \\
d_{xz} \\
d_{xy}%
\end{bmatrix}
\\
l_{z} &=&%
\begin{bmatrix}
0 & i & 0 \\
-i & 0 & 0 \\
0 & 0 & 0%
\end{bmatrix}%
\rightarrow
\begin{bmatrix}
d_{yz} \\
d_{xz} \\
d_{xy}%
\end{bmatrix}%
\end{eqnarray*}

The diagonalization of the Hamiltonian model, restricted to the local interaction terms for the Ru ion as described by the equation Eq.~\ref{eq:Hloc}, yields the complete multiplet manifolds for the $d^4$, $d^3$ and $d^5$ configurations, which are schematized in Fig. 4 c) of the main text for a representative selection of the parameters.

For the local Hamiltonian at the oxygen site $H^O_{\mathrm{loc}}$, we include only the on-site energy term to account for the energy difference between the occupied orbitals of O and Ru:
\begin{equation}
H^O_{loc}(j) =\varepsilon^{\alpha} _{x}n_{j,\alpha, px}+\varepsilon^{\alpha} _{y}
n_{j,\alpha,py}+\varepsilon^{\alpha}_{z}n_{j,\alpha, pz}\ ,
\end{equation}
where the index $\alpha$  distinguishes among planar and oxygen sites.\\
Additionally, we consider Ru-O hopping, including all symmetry-allowed terms according to the Slater-Koster rules \cite{Slater1954}, for bonds connecting Ru orbitals to apical (along the $z$ direction) and planar oxygen (along the $x$ direction). For a generic bond connecting Ru and O ions along a given symmetry direction, the $d-p$ hybridization includes all terms allowed by Slater-Koster rules and is expressed as:
\begin{eqnarray*}
H_{{Ru}-{O}} &=& t_{d_{\alpha},p_{\beta}}\left(
d_{i,\alpha\sigma }^{\dagger }p_{i+a,\beta\sigma}+H.c.\right)
\end{eqnarray*}
where the hopping $t_{d_{\alpha},p_{\beta}}$ depends on the bond angle $\theta$, and $a$ is the distance connecting the lattice sites of Ru and O. By symmetry considerations, hybridization terms in all directions can be written.

For model evaluation, we use material-specific values such as $\lambda$ in the range [0.06, 0.075] eV, $U$ [2.0, 2.2] eV, $J_H$ [0.35, 0.5] eV and $\delta$ [0.2, 0.3] eV. Similar values for $\delta$, $U$, and $J_H$ have been used for calculating electronic spectra in Ca$_2$RuO$4$, with the ratio $g = \delta / (2 \lambda)$ typically in the range $\sim$ [1.5, 2] eV for modeling spin excitations observed by neutron scattering. For the hopping amplitudes, the basic $p-d$ hopping amplitude symmetry is assumed to be $t^{0}_{p,d} = 1.5$ eV, while $\theta$ falls within the experimental range of 10\degree $-$ 15\degree ~\cite{Porter2018MagneticRuO_4}.
Hopping parameters and local orbital energies were adjusted by referencing ab initio density functional theory (DFT) 
estimate of the the nearest-neighbor Ru-Ru hopping integrals and crystal field energies
of the Ru $d$ and O $p$, both from Ref.~\cite{Gorelov2010PRL} and according to the procedure detailed in Section \ref{sec:abinitio}.\\

\subsection{Charge defects}\label{sec:charge_defects}

\begin{figure*}
\centering
\includegraphics[width=\linewidth]{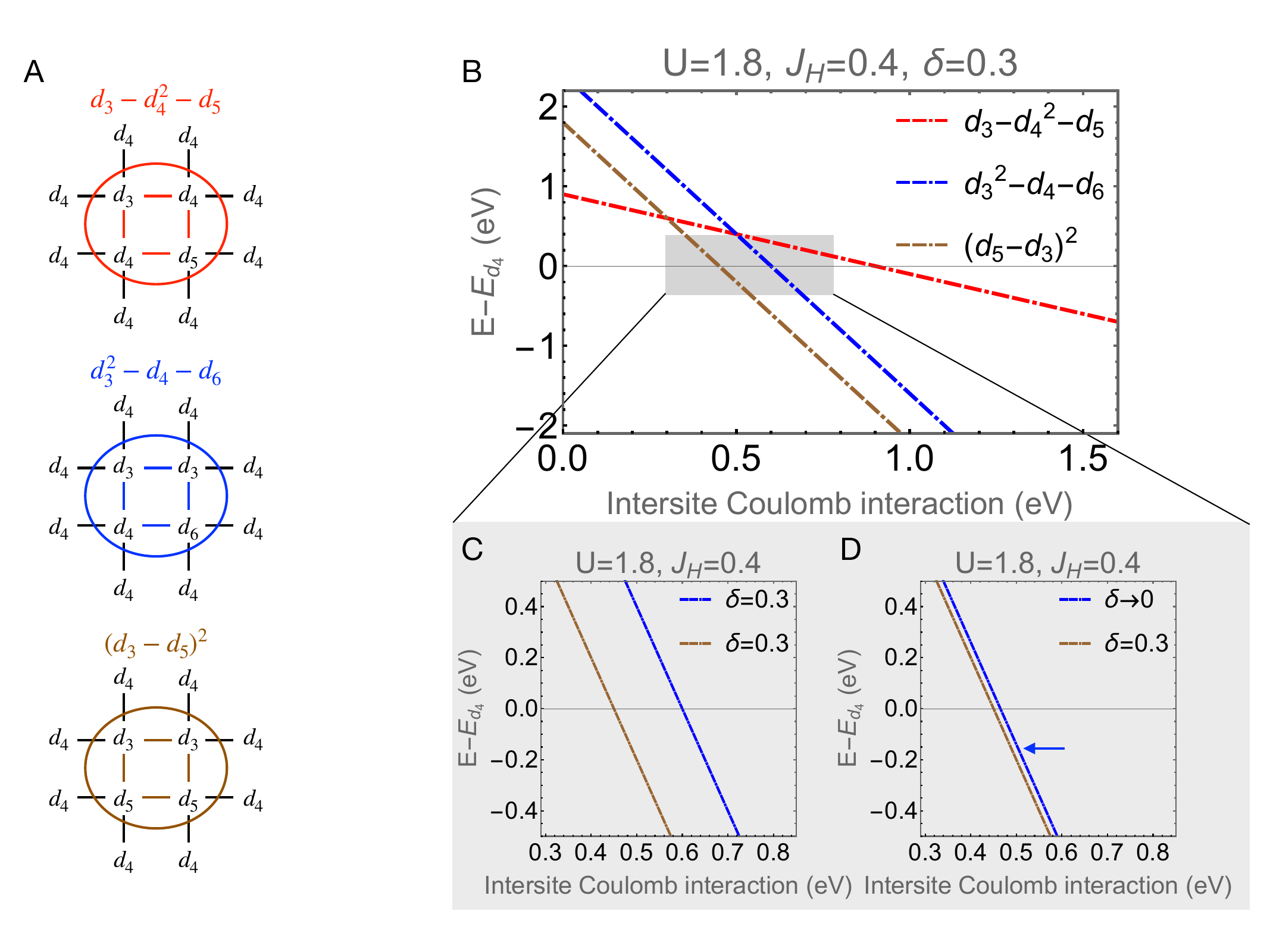}
    \caption{\textbf{Energy of charge defects on a 2$\times$2 plaquette.} A Possible charge-carrying excitations on a 2 $\times$2 plaquette with a fixed number of 16 electrons (4 electrons for each Ru site). B Energy of charge defects, computed at fixed values of the on-site Hubbard interaction $U$, the Hund's coupling $J_H$ and of the crystal field splitting $\delta$, as a function of the inter-site Coulomb interaction. C-D Zoom-in of the gray-shaded area of panel B tuning the crystal field splitting.}
    \label{fig:fig_plaq}
\end{figure*}

In this section, we provide a phenomenological 
argument to support the fact that, for suitable values of the system parameters, $d_3$, $d_5$ and $d_6$ correlated defects coexisting with $d_4$ states can become energetically favorable over the insulating ground state made solely of $d_4$ configurations. As already discussed in the main text, this result originate from a combination of octahedral distortions and modifications of the short-range Coulomb interaction and further amplified by the electric dipoles forming at the interface between short and long octahedra due to the inversion symmetry breaking \cite{van1995influence,meinders1995atomic}. To investigate this further, we examine the prototypical case of a 2$\times$2 plaquette made of 4 Ru sites, with a total of 16 electrons (4 electrons/Ru site). As illustrated in panel A of Fig.\ \ref{fig:fig_plaq}, there are three possible configurations, considering permutations: $d_3-d_4^2-d_5$, $d_3^2-d_4^2-d_6$ and $(d_3-d_5)^2$, where the square bracket notation indicates an electron configuration that appears on two sites. Furthermore, each plaquette is surrounded by 8 nearest-neighbor $d_4$ sites. The energy of each configuration is computed as the expectation value, for that configuration, of the local ruthenium Hamiltonian (Eq. \ref{eq:Hloc}) plus the short-range inter-site Coulomb interaction $\hat{H}_{
nn}=v\sum_{<i,j>}\sum_{\alpha,\beta}n_{i\alpha}n_{j\beta}$ \cite{van1995influence,meinders1995atomic}.  The lowest energy for the three configurations, by neglecting the spin-orbit coupling, is computed with respect to the one $E_{d_4}$ of the plaquette made by 4 $d_4$ sites. Thus 
\begin{eqnarray}
E_{d_3-d_4^2-d_5}-E_{d_4} &=&  3 U - 9 J_H + 2 \delta -  v, \\
E_{d_3^2-d_4-d_6}-E_{d_4} &=&  U - 3 J_H + \delta - 4v, \\
E_{(d_3-d_5)^2}-E_{d_4} &=& 2 (U - 3 J_H + \delta -2v ).
\end{eqnarray}
These three energy configurations are plotted in Fig.\ \ref{fig:fig_plaq}B as a function of the inter-site Coulomb interaction $v$, for different values of $U$, $J_H$ and $\delta$. For the parameter choices shown in Fig.\ \ref{fig:fig_plaq}B, the configurations $(d_3-d_5)^2$ and $d_3^2-d_4-d_6$ are energetically favored when, e.g, $v\sim 0.5$ eV. Considering that local distortions affect the crystal-field parameter $\delta$, we also observe that these two configurations can become degenerate, as shown in Fig.\ \ref{fig:fig_plaq}D. Such a scenario provides a possible mechanism in support of the predominant $d^3$ concentration assumed to explain the RIXS spectra. Overall, these aspects, along with the fact that $d_3$ and $d_5$ electron configurations are necessary for current to flow across the sample, justify the presence of such charge defects distributed homogeneously throughout the sample. As a final comment, we observe that $d_6$ configurations have fully occupied electronic shells and, as such, do not contribute to the O $K$-edge RIXS process, rendering these configurations inert during the resonant scattering event.

\subsection{Simulated RIXS cross section} \label{sec:RIXS_cross-section}
The RIXS intensity is governed by the Kramers-Heisenberg relation:
\begin{equation}
I(\omega,{\bf q},\epsilon,\epsilon^{'})=\sum_{f} |A_{fg}   (\omega, {\bf q},\epsilon,\epsilon^{'})|^{2} \delta(E_f+\omega_{k}-E_{g}-\omega_{k^{'}})
\end{equation}
where $\omega=\omega_{k^{'}}-\omega_{k}$ and $\bf{q}=\bf{k^{'}}-\bf{k}$ denote the energy and momentum transferred by the scattered photon, $\epsilon$ and $\epsilon^{'}$ are the polarization vectors of the incoming and outgoing light, and $E_g$ and $E_f$ are the energies of the initial and final states, respectively.

Under the dipole and fast collision approximation ~\cite{Aament2011}, the RIXS scattering amplitude $A_{fg}$  simplifies to:
\begin{equation}
A_{fg} =\frac{1}{i \Gamma} \langle f| R(\epsilon,\epsilon^{'},{\bf q}) |g\rangle\ ,
 \label{eq:amplitude}
\end{equation}
is the effective RIXS scattering operator describing two successive dipole transitions, and $\Gamma$ is the core-hole energy broadening. In O $K$-edge RIXS, the two dipole transitions create an oxygen 1$s$ core hole and an extra valence electron in a 2$p$ orbital, and vice versa. The scattering operator is expressed as:
\begin{equation}
R(\epsilon_{\nu},\epsilon_{\nu^{'}}) \propto \sum_{i,\sigma} e^{i {\bf q}  \cdot {\bf r}_i} p_{\nu^{'} \sigma}p_{\nu \sigma}  ,
 \label{eq:matrixelement}
\end{equation}
where $\nu$ stands for one of the $(x, y, z) $ orbitals and the sum over different spin states is implied. Matrix elements are then evaluated among oxygen valence states in Eq. \ref{eq:matrixelement}. Notably, the valence electron in a 2$p$ orbital hybridizes and interacts with the Ru $d$ electrons.

In the chosen experimental scattering geometry, the dependence on the incident angle $\theta_{in}$ and the scattering angle $\zeta$  between the incoming and outgoing polarization vectors is:
\begin{eqnarray}
\epsilon_{LH}&=&\epsilon_x \sin \theta_{in}+\epsilon_z \cos \theta_{in} \nonumber\\
\epsilon_{LV}&=&\epsilon_y\\
\epsilon^{'}&=&\epsilon^{'}_x \sin (\zeta-\theta_{in})+\epsilon^{'}_y+\epsilon^{'}_z \cos(\zeta-\theta_{in}) \nonumber
\end{eqnarray}
Here, the coordinate frame $(x,y,z)$ corresponds to the tetragonal axis frame $(a_{T},b_{T},c)$. Since the outgoing polarization is not resolved, the RIXS intensity is obtained by incoherently summing over all three polarization directions $(\epsilon^{'}_x ,\epsilon^{'}_y,\epsilon^{'}_z)$.\\
We then evaluated the spectral weight $W$ of each mode in the RIXS spectrum by integrating the intensity over the energy loss range where the relevant excitations \textit{A}, \textit{B}, \textit{C}, \textit{C}$^\prime$ and \textit{D} occur. The limits of integration were selected based on the energy range corresponding to the physical processes of interest, typically defined by the onset and cutoff of the excitation features in the theoretical spectrum.  Whenever the energy window contains multiple excitations, we sum all relevant contributions. These boundaries correspond to the energy ranges [0.05 $-$ 0.10] eV, [0.2 $-$ 0.6] eV, [0.6 $-$ 1.5] eV, [1.5 $-$ 3] eV, [3 $-$ 3.9] eV, respectively for excitations \textit{A}, \textit{B}, \textit{C}, \textit{C}$^\prime$, and \textit{D}.

\subsection{Wannierization of the low-energy Hamiltonian for the Ru t$_{2g}$ + e$_g$ manifold and apical O$p$ orbitals} \label{sec:ab_initio}
We carried out density functional theory (DFT) simulations using the Vienna Ab initio Simulation Package (VASP) \cite{Kresse93,Kresse96,Kresse96b}. The core and valence electrons were modeled using the projector augmented wave (PAW) method \cite{Kresse99}, with a plane-wave energy cutoff of 480 eV. The exchange-correlation effects were treated using the Perdew–Burke–Ernzerhof (PBE) \cite{Perdew08} generalized gradient approximation (GGA). We employed a $k$-point grid of 11 $\times$ 11 $\times$ 4. The total energy was minimized to less than 1 $\times$ 10$^{-5}$ eV.
The experimental lattice constant values $a$ = 5.3945 {\AA}, $b$ = 5.5999 {\AA}, $c$ = 11.7653 {\AA}~\cite{Friedt01} in the S-Pbca phase were used. Once the Bloch wave functions were obtained, the Wannier functions \cite{Marzari97,Souza01} were constructed with the WANNIER90 code \cite{Mostofi08}.
The hopping parameters were extracted in the non-magnetic phase.
This computational framework has been successfully used in the literature \cite{ma15196657,Cuono23orbital}.


\subsection{Wannerization of the t$_{2g}$ + e$_g$ manifold with apical oxygens}
\label{sec:abinitio}

We have considered the Ru atom located at (0, 0, 0) and its two apical oxygens (represented by ``a" superscript of the Hamiltonian).
The two apical oxygens are indicated with top oxygen (O$_t$) and bottom oxygen (O$_b$).
The on-site matrix $H^{on-site-a}_{d-d}$ of the $d$-orbitals is of 5$\times$5 size, considering the two-oxygen atoms the Hamiltonian $H^{on-site-a}_{p-p}$ is of 6$\times$6 size, and the d-p hybridization $H^{on-site-a}_{d-p}$ is represented by the 5$\times$6 matrix size. The total on-site Hamiltonian is 11$\times$11. Using the basis ($d_{xy}$, $d_{xz}$, $d_{yz}$, $d_{x^2-y^2}$, $d_{z^2}$, $p_{x}$-O$_t$, $p_{y}$-O$_t$, $p_{z}$-O$_t$, $p_{x}$-O$_b$, $p_{y}$-O$_b$, $p_{z}$-O$_b$), the matrices are:\\

\begin{tabular}{c|c}
\hline
 $H^{on-site-a}_{d-d} = $& $H^{on-site-a}_{d-p} = $\\
$\begin{pmatrix}
5033 & 170 & 326  & 1017 & -72\\
170 & 4307 & -60 & 394 & -683\\
326 & -60 & 4230  & 539 & 342\\
1017 & 394 & 539 & 7212 & 119\\
-72 & -683 & 342  & 119 & 6104
\end{pmatrix}$
& 
$\begin{pmatrix}
-63 & 170 & -103  & 63 & -170 & 103\\
-673 & -101 & -819 & 673 & 101 & 819\\
-104 & -812 & 528  & 104 & 812 & -528 \\
133 & 66 & 45 & -133 & -66 & -45 \\
-671 & 284 & 2083  & 671 & -284 & -2083
\end{pmatrix}$
\\
\hline
$H^{on-site-a}_{p-p} = $&\\
$\begin{pmatrix}
2709 & 14 & 107   & -321 & -8 & -112 \\
14 & 2608 & -35   & -8 & -337 & 0 \\
107 & -35 & 1872  & -112 & 0 & 148 \\
-321 & -8 & -112  & 2710 & 14   & 107 \\
-8 & -337 & 0  & 14   & 2608 & -35 \\
-112 & 0 & 148  & 107  & -35  & 1872
\end{pmatrix}$\\
\hline
\end{tabular}\\




Diagonalizing the on-site Hamiltonian and considering the energy differences we obtain 
$\delta$E$_{{\alpha=1,..6}}^{on-site}$=$\begin{pmatrix}
-4658 \\
-2617 \\
-2493 \\
-2410 \\
 -2357\\
-2243
\end{pmatrix}$
$\Delta$E$_{{\alpha=7,8,9}}^{on-site}$=$\begin{pmatrix}
0 \\
224 \\
254
\end{pmatrix}$
$\Delta$E$_{{\alpha=10,11}}^{on-site}$=$\begin{pmatrix}
3134 \\
3475
\end{pmatrix}$
with the lowest 6 eigenvalues $\alpha$=1,.., 6 being mainly \textit{p}-orbitals, the eigenvalues  $\alpha$=7, 8, 9 are mainly t$_{2g}$ and for $\alpha$=10, 11 we have the e$_g$ eigenvalues.

Regarding the terms that are not on-site, 
we report the hopping t$_{2g}$ $-$ t$_{2g}$ (7 parameters) in Tables \ref{tab:table3} and \ref{tab:table4}.
The hopping e$_g$ $-$ e$_g$ (3 parameters) are reported in Tables \ref{tab:table5} and \ref{tab:table6}. The extended model with the larger basis has parameters that are different from the restricted model, the main changes regard the hopping t$_{xy,xy}$ which is strongly suppressed while the hopping t$_{xz,xy}^{100}$ increases. Regarding the e$_g$ parameters, the largest one t$_{x^2-y^2,x^2-y^2}^{100}$ correspond to a $\sigma$-bond.

\begin{table}
  \begin{center}
    \caption{t$_{2g}$ hoppings for the model with apical oxygens from Ru in (0.5 0.5 0) to Ru in (0 0 0).  Energy units are in meV.}
    \label{tab:table3}
    \begin{tabular}{c|c|c|c|c|c|c|c|}
     lmn  & t$_{yz,yz}$ & t$_{yz,xz}$ & t$_{yz,xy}$ & t$_{xz,yz}$ & t$_{xz,xz}$ & t$_{xz,xy}$ & t$_{xy,xy}$ \\
      \hline
      100 & 23 & 50 & $-$17 & $-$51 & $-$315 & 237 & $-$20 \\ 
      010 & $-$315 & 50  & 223  & $-$51 & 23 & 5 & $-$20 \\
    \end{tabular}
  \end{center}
\end{table}

\begin{table}
  \begin{center}
    \caption{t$_{2g}$ hoppings for the model with apical oxygens from Ru in (0 0 0) to Ru in (0.5 0.5 0). Energy units are in meV.}
    \label{tab:table4}
    \begin{tabular}{c|c|c|c|c|c|c|c|}
     lmn  & t$_{yz,yz}$ & t$_{yz,xz}$ & t$_{yz,xy}$ & t$_{xz,yz}$ & t$_{xz,xz}$ & t$_{xz,xy}$ & t$_{xy,xy}$ \\
      \hline
      100 & 23 & $-$51 & 5 & 50 & $-$315 & 223 & $-$20 \\ 
      010 & $-$315 & $-$51  & 237  & 50 & 23 & $-$17 & $-$20 \\
    \end{tabular}
  \end{center}
\end{table}

\begin{table}
  \begin{center}
    \caption{e$_g$ hoppings for the model with apical oxygens from Ru in (0.5 0.5 0) to Ru in (0 0 0).  Energy units are in meV.}
    \label{tab:table5}
    \begin{tabular}{c|c|c|c|}
     lmn  & t$_{x^2-y^2,x^2-y^2}$ & t$_{x^2-y^2,z^2}$ & t$_{z^2,z^2}$ \\
      \hline
      100 & $-$206 & 155 & $-$111  \\ 
      010 & $-$206 & $-$135  & $-$111   \\
    \end{tabular}
  \end{center}
\end{table}

\begin{table}
  \begin{center}
    \caption{e$_g$ hoppings for the model with apical oxygens from Ru in (0 0 0) to Ru in (0.5 0.5 0). Energy units are in meV.}
    \label{tab:table6}
    \begin{tabular}{c|c|c|c|}
     lmn  & t$_{x^2-y^2,x^2-y^2}$ & t$_{x^2-y^2,z^2}$ & t$_{z^2,z^2}$ \\
      \hline
      100 & $-$206 & 135 & $-$111  \\ 
      010 & $-$206 & $-$155  & $-$111   \\
    \end{tabular}
  \end{center}
\end{table}

\clearpage
\bibliography{VB_Library}